\documentclass[nofootinbib,aps,11pt,letterpaper,preprintnumbers,superscriptaddress,showkeys]{revtex4}
\usepackage{amsmath}
\usepackage{eurosym}
\usepackage{amssymb}
\usepackage{graphicx}
\usepackage{color}
\usepackage{braket}
\usepackage{booktabs}
\usepackage{placeins}

\usepackage{dcolumn}
\usepackage{multirow}
\usepackage{amsmath}
\usepackage{amssymb}
 \usepackage{subfigure}
 \usepackage{caption}

\usepackage{amsfonts}
\usepackage[left=1.3cm,right=1.3cm,top=1.5cm,bottom=1.5cm]{geometry}
\usepackage[dvipsnames,svgnames]{xcolor}  
\usepackage{float}
\let\label\oldlabel
\let\ref\oldref
\usepackage[hypertexnames=false]{hyperref}
\usepackage{cleveref}   

\definecolor{darkgreen}{rgb}{0.1,0.6,0.1}
\definecolor{darkblue}{rgb}{0,0,0.3}
\definecolor{darkred}{rgb}{0.7,0,0}

\definecolor{light gray}{RGB}{220,220,220}
\definecolor{dark purple}{RGB}{108,0,217}
\definecolor{pink}{RGB}{190,20,100}
\definecolor{orang}{RGB}{193,63,0}
\definecolor{green}{RGB}{11,98,17}
\definecolor{darkpink}{RGB}{153,0,76}
\definecolor{bluegreen}{RGB}{0,102,102}
\definecolor{greenlagan}{RGB}{0,102,0}
\definecolor{redgreen}{RGB}{102,102,0}
\definecolor{Redgreen}{RGB}{153,76,0}
\definecolor{vividviolet}{rgb}{0.62, 0.0, 1.0}
\definecolor{amaranth}{rgb}{0.9, 0.17, 0.31}
\definecolor{palatinateblue}{rgb}{0.15, 0.23, 0.89}
\definecolor{brightpink}{rgb}{1.0, 0.0, 0.5}
\definecolor{cornflowerblue}{rgb}{0.39, 0.58, 0.93}
\definecolor{deepcarminepink}{rgb}{0.94, 0.19, 0.22}
\definecolor{radicalred}{rgb}{1.0, 0.21, 0.37}
 
\definecolor{beamer@PRD}{RGB}{46,48,146}
\hypersetup{
     breaklinks=true,
    pdfstartview={FitH},    
    colorlinks=true,       
    linkcolor=blue,          
    citecolor=cyan,        
    filecolor=magenta,      
    urlcolor=magenta,           
    anchorcolor=cyan,      
    linktocpage=true
}

\begin{document}
	{\vskip .1cm}
\date{\today}
\newcommand\be{\begin{equation}}
\newcommand\ee{\end{equation}}
\newcommand\bea{\begin{eqnarray}}
\newcommand\eea{\end{eqnarray}}
\newcommand\bseq{\begin{subequations}} 
\newcommand\eseq{\end{subequations}}
\newcommand\bcas{\begin{cases}}
\newcommand\ecas{\end{cases}}
\newcommand{\p}{\partial}
\newcommand{\f}{\frac}

\title{\Large Shadow curves and quasinormal modes for rotating black holes
surrounded by dark matter, radiation and dust\normalsize}

\author {\textbf{Reza Pourkhodabakhshi}}\email{reza.pk.bakhshi@gmail.com
}
\affiliation{ Departament de F\' \i sica Cu\' antica i Astrof\'\i sica and Institut de Ci\`encies del Cosmos, 
Universitat de Barcelona, Mart\'i Franqu\`es, 1, 08028
Barcelona, Spain.}
\author {\textbf{ Jorge G.~Russo}}
\email{jorge.russo@icrea.cat
}
\affiliation{ Departament de F\' \i sica Cu\' antica i Astrof\'\i sica and Institut de Ci\`encies del Cosmos, 
Universitat de Barcelona, Mart\'i Franqu\`es, 1, 08028
Barcelona, Spain.}
\affiliation{ Instituci\'o Catalana de Recerca i Estudis Avan\c{c}ats (ICREA),\\
Pg. Lluis Companys, 23, 08010 Barcelona, Spain.}


\begin{abstract}

We study  the impact of different fluid matter scenarios on the shadow of black holes (BH) and on frequencies of quasinormal modes (QNM)   
for a black hole subjected to scalar perturbations, with a comparison to the standard Kerr. 
The analysis of the shadow reveals a dependence on the density parameter of  fluid matter $k$, with larger shadow sizes for larger values of $k$ in the dark matter and dust cases, while the shadow size becomes smaller in the radiation case.
Notably, dark matter induces  more visible deformations compared to radiation or dust, thereby highlighting its distinct imprint on shadow curves. 
We also find that dark matter reduces the real part of the QNM frequencies and significantly increases the damping time, enhancing the prospects for gravitational wave detection.
The variations in  spin parameter $a$, density parameter $k$ and multipole number $l$ are investigated.
The analysis confirms  the significant role of dark matter in modifying the behavior of QNMs, providing a promising avenue for future experiments.

\bigskip


 \end{abstract}

\maketitle

\section{Introduction}

For the first time, the existence of BHs as real objects in our universe was confirmed by the release of a direct image of the supermassive BH M87* by the Event Horizon Telescope (EHT) \cite{EHT}. 
When light passes close to a BH, the rays can be deflected significantly and may even travel in circular orbits. This strong deflection, combined with the fact that no light escapes from a BH, results in a BH appearing as a dark disk in the sky, known as the BH shadow. The idea of observing this shadow was discussed by \cite{Falcke_2000}, focusing on the supermassive BH at the center of our Galaxy, associated with the radio source Sagittarius $A^*$. EHT enables some of the first horizon-scale tests of gravity, placing tight constraints on deviations from General Relativity and black hole mimickers \cite{Vagnozzi_2023}.

Synge \cite{Synge} was the first to calculate what we now refer to as the shadow of a Schwarzschild black hole. The first numerical calculation of the visual appearance of a Schwarzschild BH surrounded by a shining and rotating accretion disk was conducted by Luminet \cite{Luminet}, who assumed that light travels on lightlike geodesics of the space-time metric. Viergutz \cite{Viergutz} extended this work to the Kerr space-time. For an overview of such numerical studies, which are not the focus of this article, we refer to James et al. \cite{James2015}. The observational appearance of a black hole, as obtained in numerical simulations, strongly depends on the distribution of light sources and the properties of the emitting matter near the black hole \cite{Broderick_2011,10.1111/j.1365-2966.2005.09458.x,Broderick_2014,Bronzwaer_2021,Dexter_2009,Kardashev_2014,PhysRevD.100.024018,Narayan_2019}. While resulting images can vary greatly, the main feature—the shadow—will consistently have the same size and shape, determined by the propagation of light in the strong gravitational field of the black hole. Determining the visual appearance of the shining matter, most likely an accretion disk, is the subject of ongoing numerical studies.
In our analysis, we utilized methods similar to those found in \cite{PERLICK2022,Bardeen,Cunha,Grenzebach,Teo}
(see also \cite{Tsukamoto_2014,Tsukamoto_2018} for useful analytic formulas).

With the development of gravitational-wave (GW) astronomy \cite{LIGO}, we are now uniquely positioned to test Einstein's theory of general relativity (GR) at unprecedented levels \cite{Berti_2015,Berti46B,Berti49B,Pretorius}. One of the most striking predictions of GR is that the unique stationary and asymptotically flat black hole (BH) solution is described by the Kerr geometry \cite{Carter,Israel}. This significant fact implies that, to a very good approximation, all BHs in the universe are expected to be typically uncharged and uniquely described by their mass and spin, a result commonly referred to as the ``no-hair hypothesis" \cite{Cardoso_2016}. Similar no-hair theorems have also been established in the context of modified gravity, notably in some classes of scalar-tensor theories \cite{Radu.Eugen}. However, BHs do exhibit ``hair" in several extended theories \cite{Babichev_2015,Khodadi222_2021}, though recent results imply that such hair must extend beyond the innermost light ring \cite{Ghosh_2023}.

Testing the validity of the no-hair hypothesis is one of the most exciting prospects for future GW detectors \cite{Berti49B,Robinson,Cardoso_2016,Carter,Israel}. When slightly perturbed, BHs relax back to equilibrium through the emission of GWs characterized by a set of exponentially damped sinusoids with specific frequencies and damping times, known as quasinormal modes (QNMs), which dominate the late stages of the GW signal emitted from a binary BH merger \cite{Berti49B,Robinson,Cardoso_2016,Carter,Israel,Sotiriou_2015,Starinets}. For a Kerr BH in GR, the entire QNM spectrum is uniquely determined by the BH's spin and mass. Therefore, detecting multiple QNM frequencies from the remnant of a BH merger would enable us to conduct non-trivial tests of the no-hair hypothesis and potentially identify deviations from the Kerr geometry \cite{Olaf,Alessandra,Veitch,Clifford,Meidam}.

These experimental prospects necessitate a theoretical effort to compute the QNMs of BHs in alternative theories of gravity. This can be done in a parametrized theory-agnostic fashion \cite{Ferreira,Kostas} or case by case for specific examples of modified theories of gravity. Notably, QNM frequencies for BHs in theories beyond GR have been computed for spherically symmetric solutions in theories such as Einstein-Maxwell-Dilaton \cite{Konoplya_2002,Massimo}, Einstein-Dilaton-Gauss-Bonnet gravity \cite{Gualtieri}, and dynamical Chern-Simons gravity \cite{Pani}. The extension to spinning BHs, beyond the known QNMs of Kerr BHs in GR, has been primarily hindered by the lack of known exact analytical solutions in modified theories of gravity and the general difficulty of separating the equations of motion describing perturbations in those theories when dealing with axisymmetric spacetimes. The notable exception to this rule is the Kerr-Newman case in Einstein-Maxwell theory, for which QNMs have only recently been computed \cite{Godazgar,Leonardo}.

Our primary method—and the second significant simplification we employ—is the utilization of the eikonal (or geometric optics) approximation to derive QNM solutions from the wave equations. This approximation has a rich and successful history in the examination of Schwarzschild and Kerr black holes, dating back to the early 1970s \cite{Goebel,Press} (for comprehensive reviews, see \cite{Starinets,Kokkotas_1999}). According to the established eikonal framework, the fundamental QNM can be envisioned as a wave packet concentrated in the radial direction at the peak of the wave potential; in this approximation, the peak itself aligns with the location of the photon ring of null geodesics. 
This property extends to photon surfaces, which, under energy conditions, must enclose black holes in arbitrary spacetime \cite{Claudel_2001}.
The real part of the QNM frequency is determined to be an integer multiple of the orbital angular frequency at the photon ring. Similarly, the decay rate (imaginary part) of the QNM is correlated with the Lyapunov exponent of the unstable null orbits at the photon ring radius \cite{Mashhoon,Bahram,Witek}. This intuitive eikonal model has also been employed to establish a relationship between the $\ell>|m|$ QNMs of Kerr black holes (where $\ell$ and $m$ are the usual spherical harmonic integers) and the nonequatorial spherical photon orbits \cite{Dolan,Nichols}. More recently, a post-Kerr parametrized scheme inspired by the QNM-photon ring relation has been developed as a model for the fundamental $\ell=|m|$ mode of non-GR black holes \cite{Kostas}. While this model can describe the more astrophysically relevant case of rotating black holes, it does not consider any additional field degrees of freedom. QNMs of rotating black holes exhibit a correspondence with their shadow radii, extending beyond the Kerr metric to more general rotating spacetimes \cite{Pedrotti_2024}.
Our approach is similar to that of Ref. \cite{Dolan} for Kerr black holes.

Any endeavor aimed at exploring the genuine essence of black holes must accommodate deviations from GR's Kerr spacetime, as well as considering theoretical insights from alternative theories of gravity. One of the simplest strategies beyond Kerr is to employ parametrized schemes, where parameters regulate the deviation from GR, for both the black hole's spacetime metric \cite{Johannsen,Tim,Rezzolla} and the associated QNMs \cite{Masashi,Kostas}, without committing to any specific theory of gravity. However, this approach's primary limitation is that it may merely serve as a non-trivial test of the Kerr metric, as the deformations might not correspond to any particular gravity theory. The alternative, more rigorous (albeit considerably more demanding) strategy involves theoretically computing black hole spacetimes and their GW signatures on a case-by-case basis within the array of modified gravity theories. Notably, this latter approach is substantially more challenging to execute, leading to QNMs of non-GR black holes being computed only for a limited number of cases, typically assuming spherical symmetry \cite{Pani,Gualtieri,Brito} (for a comprehensive review and additional references,  see \cite{Berti_2015}).

Astrophysical BHs are not isolated from matter, 
therefore it is important to study the effect of matter around the $\mathrm{BH}$ on the shadow and on the QNMs as well. Such a study is well-justified, since, in the framework of scalar-tensor gravity, it has been shown the matter around $\mathrm{BH}$, significantly affects the massive scalar superradiance of the rotating $\mathrm{BH}$ \cite{Cardoso_2013-1,Cardoso_2013-2}. Possible scenarios include black holes surrounded by various types of matter in galactic halos, such as dark matter halos, anisotropic fluids, or black holes in an expanding universe\cite{Li_2020,Konoplya_2022,Konoplya_2019,Lee_2021,chakraborty2025tidallovenumbersquasinormal,Khodadi2222_2022,Chowdhuri_2021,Adler_2022}.

While presently it is not  known what kind of matter dominates the region around the $\mathrm{BH}$, one can parametrize
different types of matter by using the Kiselev spacetime \cite{Kiselev_2003,toshmatov2015rotating}. This describes a family of exact solutions in which the rotating $\mathrm{BH}$ is surrounded by three types of anisotropic fluid matter; radiation, dust, dark matter with Equation-of-States (EoSs) $\alpha=1 / 3,0$ and $-1 / 3$, respectively.
Recently, for the charged Kiselev BH surrounded by these three types of anisotropic fluid matter, superradiance and instability were analyzed \cite{Cuadros_Melgar_2021}.

The study of QNMs based on the gravitational perturbations is well-motivated phenomenologically, where gravitational perturbations consist of spin $0$, $1$ and $2$ components \cite{Kostas,Masashi}. Here we will focus on the spin-$0$ component of gravitational perturbations. 

In Sec. \ref{sec:Kiselev-analysis} we introduce the Kiselev spacetime and study the stress-energy tensor.
According to the values of the parameters, this can describe
different types of fluid matter. In particular, conformal matter (radiation), dust  or dark matter.
We also study constraints on parameters from the weak energy condition (WEC).

In Sec. \ref{ch:BH-shadow} we calculate
and provide a thorough  analysis of the shadow of a rotating Kiselev black hole, representing
a rotating black hole surrounded by fluid matter.
Following  \cite{PERLICK2022}, we will first establish a general formula for the inclination angle of a light ray that an observer emits into the past. This formulation remains broadly applicable due to the inclusion of an unspecified constant of motion.
 Further details on the method are described in   the appendix (see also \cite{PERLICK2022}).

In Sec. \ref{ch:QNM}, we delve into the concept of quasinormal modes (QNMs) and elucidate the derivation of the perturbation equation,  known as the Regge-Wheeler equation \cite{Wheeler}, originating from scalar metric perturbations. Furthermore, we delve into the practicalities of calculating QNMs, introducing a Schrodinger-like wave equation in terms of an effective potential \cite{Chandrasekhar:1985kt,Detweiler,PANI_2013,Zerilli}. We will also make use of the eikonal approximation \cite{Glampedakis}, which gives  further insight into the structure of QNMs.

In order to study the QNMs in the rotating Kiselev spacetime, here we will use the Newman-Penrose formalism \cite{Penrose,Penrose_Rindler_1984} to first derive the axial perturbation equation for a scalar field, the Teukolsky equation \cite{Teukolsky}. This fundamental equation is separated into radial and angular components. Specifically, we focus on solving the radial equation utilizing the eikonal approximation method \cite{Glampedakis}. 


\section{The Rotating Kiselev Metric}\label{sec:Kiselev-analysis}

By applying the Newman-Janis algorithm \cite{osti_4608715} in the conventional manner to derive a rotational spacetime metric from its spherically symmetric counterpart, and incorporating the suggested modifications in \cite{Azreg_A_nou_2014}, the solution for a rotating Kiselev BH in Boyer-Lindquist coordinates \((t, r, \theta, \varphi)\) can be expressed as follows \cite{Rahaman_2010}:

\begin{equation}
\begin{aligned}
ds^2 = &-\left(\frac{\Delta_k - a^2 \sin^2 \theta}{\Sigma}\right) dt^2 + \frac{\Sigma}{\Delta_k} dr^2
- 2a \sin^2 \theta \left(\frac{\Delta_k - (r^2 + a^2)}{\Sigma}\right) dtd\varphi \\
&+ \Sigma d\theta^2  + \sin^2 \theta \left(\frac{(r^2 + a^2)^2 - a^2 \Delta_k \sin^2 \theta}{\Sigma}\right) d\varphi^2,
\end{aligned}
\label{eq:Kiselev-Metric-QNM}
\end{equation}
where
\begin{equation}
\Sigma = r^2 + a^2 \cos^2 \theta, \quad \Delta_k = r^2 - 2mr + a^2 - kr^{1 - 3\alpha}.
\label{eq:delta-Kiselev}
\end{equation}
Here, $m$ and $a$ are the BH mass and rotation parameter of the BH. $k$ is an integration constant controlling the amount of fluid matter and that distinguishes the above metric from the standard Kerr metric $(k = 0)$. $k$ can be expressed in terms of a dimensionless parameter $\kappa$ defined by 
$k=\kappa m^{1+3\alpha}$.
Finally, $\alpha$ is the equation of state (EoS) parameter that determines the type of dominant matter distribution around the BH. In the real universe, these three types of matter may be mixed. However, for simplicity, we assume that the EoS considered here represents just the dominant effect of different species  around the BH. This simplification allows us to model the complex interactions in a more manageable way by focusing on the primary contribution of each type of matter to the overall metric.
Thus, while multiple forms of matter could coexist around a BH, our assumption isolates the dominant species for clearer analysis. This approach aids in understanding the individual influences of radiation $(\alpha = \frac{1}{3})$, dust $(\alpha = 0)$, and dark matter $(\alpha = -\frac{1}{3})$ on the spacetime geometry and on the black hole shadow. Each type of matter imparts distinct characteristics to the metric, which are essential for our investigation.
 While this is an approximation, it provides valuable insights into the gravitational effects exerted by different forms of matter around a rotating BH.

\bigskip

It is noteworthy that dark matter is generally modeled in various formats; here, it is considered as a fluid matter \cite{Rahaman_2010,Xu_2018}. An interesting feature of the Kiselev metric is its ability to recover some well-known classical solutions depending on the type of fluid matter considered. For $\alpha = \frac{1}{3}$, the metric effectively describes a Kerr BH enclosed by radiation; indeed, the geometry is equivalent to the Kerr-Newman BH with $-k$ being proportional to the squared electric charge.  For $\alpha = 0$, it represents a Kerr BH with a shifted mass. 

For these three types of matter, the metric function $\Delta_k$ has two real roots, $r_{\pm}$, where the larger root $r_{+}$ denotes the location of the event horizon (EH). Two important points must be noted regarding the metric \eqref{eq:Kiselev-Metric-QNM}. Firstly, despite a common misconception in the literature, the metric \eqref{eq:Kiselev-Metric-QNM} does not represent a perfect fluid but rather some type of anisotropic fluid matter. For a detailed investigation of these points, it is recommended to refer to Visser's critical paper \cite{Visser_2020}. Throughout this paper, we will refer to the metric \eqref{eq:Kiselev-Metric-QNM} as the Kiselev rotating BH including the anisotropic fluid matter.

\subsection{The stress tensor} 

 Let us first consider  the non-rotating  Kiselev metric \eqref{eq:Kiselev-Metric-QNM} with $a=0$.
 The stress-energy tensor can be computed as
 usual from Einstein field equations,  $T_{ij}\equiv 8\pi \mathcal{T}_{ij} =R_{ij}-\frac{1}{2}g_{ij}R $. This gives the components
 \begin{equation}
     \begin{aligned}
         &T_{tt}= 3 \alpha  k r^{-6 \alpha -4} \left(k+(2 m-r) r^{3 \alpha }\right),\\
         &T_{rr}= -\frac{3 \alpha  k}{r^2 \left(k+(2 m-r) r^{3 \alpha }\right)},\\
         &T_{\theta \theta} = -\frac{3 \alpha}{2}  (3 \alpha +1) k r^{-3 \alpha -1},\\
        &T_{\varphi \varphi}= -\frac{3 \alpha}{2}  (3 \alpha +1) k \sin ^2(\theta ) r^{-3 \alpha -1},
     \end{aligned}
     \label{eq:Tij-a0}
 \end{equation}
 where $T_{ij} = 0$ for $i\neq j$ .

 For the case of  $\alpha=-\frac{1}{3}$ from \eqref{eq:Tij-a0} we can calculate  $\rho = \frac{T_{tt}}{-g_{tt}} = \frac{ k }{r^2}$, $P_{rr} = \frac{T_{rr}}{g_{rr}} =-\frac{ k }{r^2}$ and $P_{\theta \theta} = P_{\varphi \varphi} = 0$. It implies that $\rho =- P_{rr} \sim \frac{1}{r^2}$ which is a characteristic of dark matter components in the stress-energy tensor, initially found via studying galactic halos \cite{Kiselev_2003,Matos_2000}. The fact $P_{\theta \theta} = P_{\varphi \varphi} = 0$ also shows the fluid matter is not a perfect fluid, but anisotropic. It is worth noting that in this case the asymptotic of the metric at infinity is not Minkowski, because the diagonal term does not vanish. However, the metric is supposed to apply only in the halo region where dark matter is supposed to be concentrated.

 Let us now incorporate the spin parameter $a$.
 The trace of stress-energy tensor $\operatorname{tr}(T_{ij})=T_{ij} g^{ij}$ is

 \begin{equation}
   \operatorname{tr}(T_{ij})  = -\frac{6 \alpha  (3 \alpha -1) k r^{-3 \alpha -1}}{a^2 \cos (2 \theta )+a^2+2 r^2}.
 \label{eq:trace-T}
 \end{equation}

For the case of $\alpha=\frac{1}{3}$, from \eqref{eq:trace-T} we get $\operatorname{tr}(T_{ij}) = 0$. The traceless stress-energy tensor implies that the case of $\alpha=\frac{1}{3}$ represents a conformally invariant matter fluid; a feature of  Maxwell theory representing radiation.
This is expected, since the metric in this case corresponds to Kerr-Newman.

For the rotating case of \eqref{eq:Kiselev-Metric-QNM}, the tt-component of stress-energy tensor is given by

\begin{equation}
    \begin{aligned}
        &T_{tt} = 3 \alpha  k \resizebox{!}{20pt}{[} r^{3 \alpha } \left((3 \alpha -1) a^4 \cos (4 \theta )\right.+(1-3 \alpha ) a^4+4 (3 \alpha +1) a^2 r^2 \cos (2 \theta )\\
        &\left.-4 (3 \alpha +5) a^2 r^2+16 r^3 (2 m-r)\right)+16 k r^3 \resizebox{!}{20pt}{]} \times \left( 2 r^{6 \alpha +1} \left(a^2 \cos (2 \theta )+a^2+2 r^2\right)^3 \right)^{-1}.
    \end{aligned}
    \label{eq:T00-rotating}
\end{equation}
 For  $\alpha=\frac{1}{3}, \ -\frac{1}{3},\  0$, we obtain
 
 \begin{equation}
 \begin{aligned}
     & T^{\textbf{Rad}}_{tt} = \frac{4 k \left(a^2 \cos (2 \theta )-3 a^2+2 k+4 m r-2 r^2\right)}{\left(a^2 \cos (2 \theta )+a^2+2 r^2\right)^3} ,\\
&T^{\textbf{Rad}}_{tt} = \frac{ k}{\Sigma^3} 
\left(\Sigma-\Delta_k-r^2-a^2\right) ,\\
          & T^{\textbf{DM}}_{tt} = \frac{k \left(a^4 \cos (4 \theta )+8 a^2 r^2+8 r^3 (r-k r-2 m)\right)}{\left(a^2 \cos (2 \theta )+a^2+2 r^2\right)^3} ,\\
           & T^{\textbf{DM}}_{tt} = \frac{k}{\Sigma^3}\left(\big(\Sigma-r^2)^2+r^2\Delta_k^2+a^2)\right) ,\\
     & T^{\textbf{Dust}}_{tt} = 0 ,
     \end{aligned}
     \label{eq:T00-alpha-cases}
 \end{equation}
where $T^{\textbf{Dust}}_{tt} = 0$ is a consequence that the dust case $\alpha=0$ is a vacuum solution of general relativity, since it corresponds to the standard Kerr metric with a shift in mass term as $2m\to 2m+k$. We consider this case to contrast the behavior of the presence of dark matter with respect to ordinary dust matter.

 The asymptotic behavior of $T_{tt}$ from \eqref{eq:T00-alpha-cases} at large $r$ is 
 as follows 
 \begin{equation}
     \begin{aligned}
         &  \alpha>-\frac13:\ \quad  T_{tt}\approx -\frac{3\alpha k}{r^{3\alpha+3}}\ 
         , \\
         & \alpha=-\frac13:\quad T_{tt} \approx \frac{k(1-k) }{r^2}\ ,\\
              &  \alpha<-\frac13:\ \quad  T_{tt}\approx \frac{3\alpha k^2}{r^{6\alpha+4}}\ .
          \end{aligned}
     \label{asymNew}
 \end{equation}
 In particular,
 \begin{equation}
     \begin{aligned}
           \alpha=\frac13:&\ \quad  T^{\textbf{Rad}}_{tt}\approx -\frac{k}{r^4}
           \\
          \alpha=-\frac13:&\quad T^{\textbf{DM}}_{tt} \approx  \frac{k(1-k) }{r^2}\ .
     \end{aligned}
     \label{eq:T00-asymptotic}
 \end{equation}
The weak energy condition $T_{tt}\geq 0$ thus requires that $k\leq 0$ for $\alpha>0$. This includes the radiation case with
$\alpha=\frac13$.
For the $\alpha=-1/3$ case one must restrict to  $0\leq k<1$ 
On the other hand, WEC is violated for theories with
$\alpha<-\frac13$.
Thus, from \eqref{eq:T00-asymptotic} it follows that the  weak energy condition $T_{tt} \geq 0 $ is satisfied provided  $k^{\textbf{Rad}}\leq0$ and $0\leq k^\textbf{DM} <1$. Using \eqref{eq:T00-alpha-cases}, one can check that these conditions on $k$ are sufficient to satisfy  the weak energy condition not only asymptotically but also in the whole spacetime outside the horizon.


\bigskip

The remaining, non-vanishing components of the stress tensor are given by
\begin{eqnarray}
    T_{rr}&=&\frac{3\alpha k r^{1-3\alpha}}{\Delta_k \Sigma}
    \nonumber\\
     T_{\theta\theta}&=& \frac32 \alpha k r^{-1-3\alpha}\left(1-3\alpha -\frac{2r^2}{\Sigma}\right)
     \nonumber\\
         T_{t\varphi}&=& \frac{3\alpha k a r^{-1-3\alpha}}{2\Sigma^3}\sin^2\theta\left( (1-3\alpha)(r^2+a^2)\Sigma-2r^2(r^2+\Delta_k^2+a^2)\right)
     \nonumber\\
    T_{\varphi\varphi}&=&\frac{3\alpha k  r^{-1-3\alpha}}{2\Sigma^3}\sin^2\theta \left[ \Sigma \big(2r^2\Delta_k+(1-3\alpha)(r^2+a^2)^2\big) -2r^2(r^2+a^2)(r^2+\Delta_k^2+a^2)\right]
    \label{segunda}
 \end{eqnarray}
The stress tensor satisfies 
\begin{equation}
    R_{ij}-\frac{1}{2}g_{ij}R=T_{ij}\ ,\qquad D_iT^{i}_{\ j}=0\ .
\end{equation}
Summarizing,  the rotating Kiselev metric is an exact solution of the Einstein equations sourced by a rotating anisotropic fluid with stress tensor components
\eqref{eq:T00-rotating}, \eqref{segunda}. By construction, the stress tensor satisfies the conservation equations, as it can be checked by direct calculation.
In addition, it obeys the weak energy condition provided $k\leq 0$ for $\alpha=\frac13$ (radiation) and $0\leq k<1$ for $\alpha=-\frac13$ (dark matter).


\subsection{Extremality in Rotating Kiselev Metric}

The horizons of the black hole are determined by the roots of the equation
\begin{equation}
    \Delta_k = 0,
    \label{eq:delta_zero}
\end{equation}
where the function \(\Delta_k\) is given by \eqref{eq:delta-Kiselev}. Extremality occurs when there is a degenerate horizon.
For the values of $\alpha$ considered in this paper, \(\alpha = -\frac{1}{3}, 0 \;\text{and}\; \frac{1}{3}\),
the locations of Cauchy and event horizons are
\begin{equation}
    \begin{aligned}
        \textbf{DM:} \quad & (1 - k)r^2 - 2 m r + a^2 =0,  \\
        & r^{\text{DM}}_\text{H} = \frac{m \pm \sqrt{m^2 - (1-k)a^2}}{(1-k)}. \\
        \textbf{Dust:} \quad & r^2 - 2m r + a^2 - k r = 0, \\
        & r^{\text{Dust}}_\text{H} = m + \frac{k}{2} \pm \sqrt{\Big(m + \frac{k}{2}\Big)^2 - a^2}. \\
        \textbf{Rad:} \quad & r^2 - 2m r + a^2 - k = 0, \\
        & r^{\text{Rad}}_\text{H} = m \pm \sqrt{m^2 - (a^2 - k)}.
    \end{aligned}
    \label{eq:horizon_equations}
\end{equation}
Extremality occurs when the discriminants in \eqref{eq:horizon_equations}  vanish.
The extremal value of the spin parameter is, in each case,
\begin{equation}
    \begin{aligned}
        \textbf{DM:}  \quad
        & a^{\text{DM}}_\text{Ext} = \frac{m}{\sqrt{1-k}}. \\
        \textbf{Dust:} \quad 
        & a^{\text{Dust}}_\text{Ext} = m + \frac{k}{2}. \\
        \textbf{Rad:} \quad 
        & a^{\text{Rad}}_\text{Ext} = \sqrt{m^2 - |k|}.
    \end{aligned}
    \label{eq:extremality_conditions}
\end{equation}
%
%
%
%
The different cases may be compared with the extremal $a=m$ condition for the Kerr black hole.
For dark matter ($\alpha=-1/3$) extremality occurs at a spin parameter  $a^\text{DM}_{\text{Ext}} > m$ for any $k>0$. The same is true for the case of dust with $k>0$ (in this case, this is obvious, since the $\alpha=0$ solution is just the Kerr solution with a shifted mass). In the case of radiation ($\alpha=1/3$) we note that extremality occurs for $a^\text{Rad}_\text{Ext} < m$ for $k< 0$ (we recall that $k\leq 0$ is
 a requirement in order for the weak energy condition to be satisfied).

\bigskip

\section{Rotating Kiselev BH Shadow}
\label{ch:BH-shadow}

In this section we will discuss the main features of
the shadows of Kiselev black holes, using the cases
$\alpha=1/3,\ 0,\ -1/3$ as illustrative examples.
The relevant formulas describing the black hole shadow are derived in full detail in the Appendix \ref{sec:rotating-BH}.
In the observer's sphere, the shadow is represented
by  azimuthal and colatitude angles $(\psi,\zeta)$, given in \eqref{eq:sin-psi-3} and \eqref{eq:sin-zeta}.
%
Compared to the shadow of the Kerr black hole, the change in the Kiselev spacetime \eqref{eq:Kiselev-Metric-QNM}  appears as a different $\Delta_k$ function \eqref{eq:delta-Kiselev}.

\begin{figure*}[t]
    \centering
    \begin{subfigure}{ }
        \centering
        \begin{minipage}{0.25\textwidth}
            \centering
            \includegraphics[width=\textwidth]{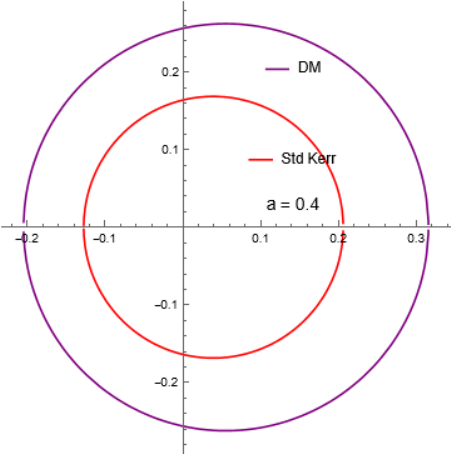}
        \end{minipage}
        \hspace{15pt}
        \begin{minipage}{0.25\textwidth}
            \centering
            \includegraphics[width=\textwidth]{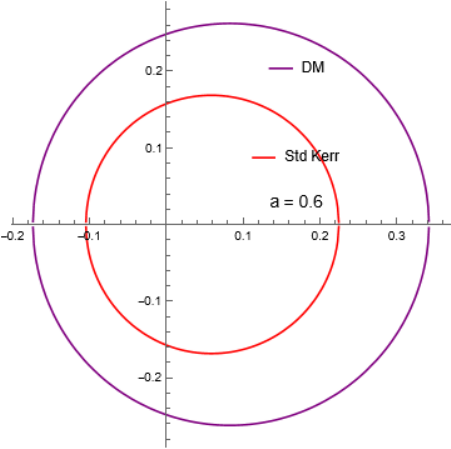}
        \end{minipage}
                \hspace{15pt}
        \begin{minipage}{0.25\textwidth}
            \centering
            \includegraphics[width=\textwidth]{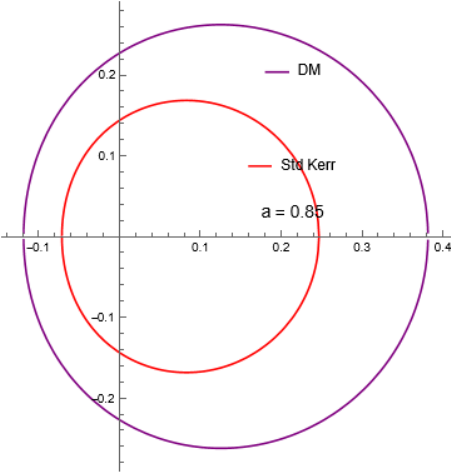}
        \end{minipage}
        \caption{Shadows for the case $\alpha=-\frac13$ compared to Kerr, for different rotation parameters,  $a=0.4 m$,  $a=0.6m$ and   $a=0.85m$, respectively from left to right ($m=1$, $\kappa=  0.2 $, $r_0= 30$). Observer is equatorial, $\theta_0=\frac{\pi}{2}$.}
        \label{fig:Dark-Matter-plots-2}
    \end{subfigure}
    \begin{subfigure}{ }
        \centering
        \begin{minipage}{0.25\textwidth}
            \centering
            \includegraphics[width=\textwidth]{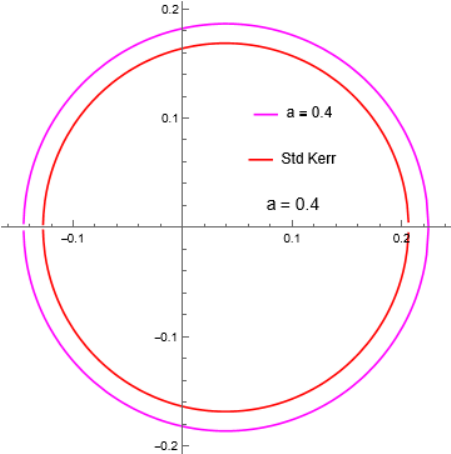}
        \end{minipage}
        \hspace{15pt}
        \begin{minipage}{0.25\textwidth}
            \centering
            \includegraphics[width=\linewidth]{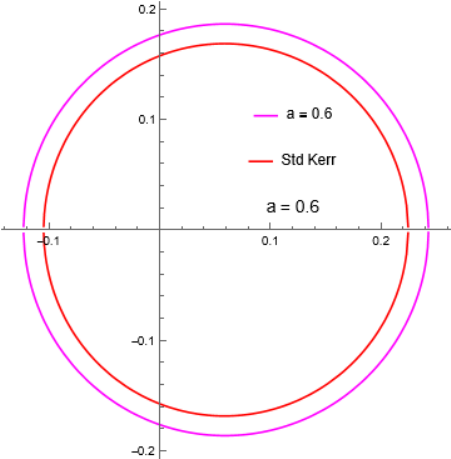}
        \end{minipage}
        \hspace{15pt}
        \begin{minipage}{0.25\textwidth}
            \centering
            \includegraphics[width=\linewidth]{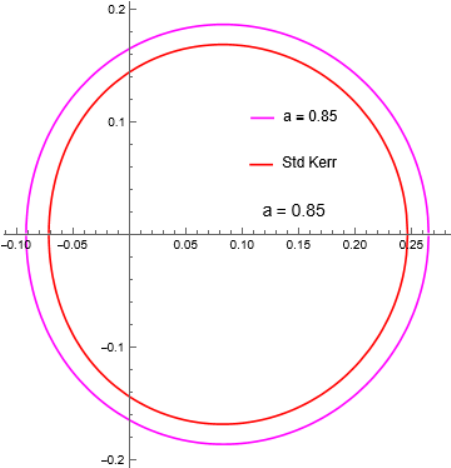}
        \end{minipage}
        \caption{Shadows for the case $\alpha=0$ compared to Kerr. Same values of parameters as in fig. 1.}
        \label{fig:Dust-plots-2}
    \end{subfigure}
    \begin{subfigure}{ }
        \centering
        \begin{minipage}{0.25\textwidth}
            \centering
            \includegraphics[width=\linewidth]{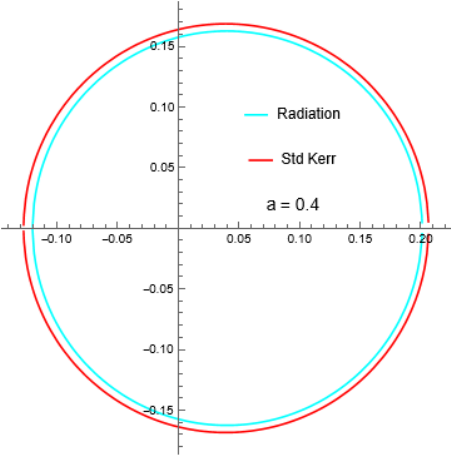}
        \end{minipage}
        \hspace{15pt}
        \begin{minipage}{0.25\textwidth}
            \centering
            \includegraphics[width=\linewidth]{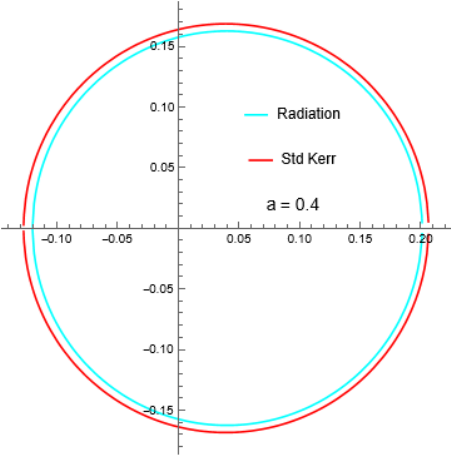}
        \end{minipage}
        \hspace{15pt}
        \begin{minipage}{0.25\textwidth}
            \centering
            \includegraphics[width=\linewidth]{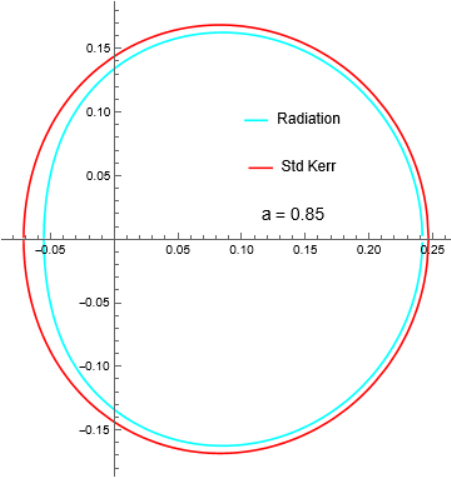}
        \end{minipage}
        \caption{Shadows for the case $\alpha=\frac13$ compared to Kerr. Here $\kappa= - 0.2$, the other parameters as in fig. 1.}
        \label{fig:rad-plots-1}
    \end{subfigure}
\end{figure*}

The photon region is determined by  \eqref{L_E-result} and \eqref{eq:K_E-final-result}.  We have

\begin{equation}
    \bar K_E^k=\frac{16 \Delta_k   r^2}{\left(\frac{\partial \Delta_k }{\partial r}\right)^2}
   \ ,\quad 
    L_E^k=\frac{1}{a}\left[\left(a^2+r^2\right)-\frac{4 \Delta_k   r}{\frac{\partial \Delta_k }{\partial r}}\right] ,
    \label{eq:Kiselev-L_E}
\end{equation}
where $\bar K_E = K_E+\left(L_E-a\right)^2$, with $L_E^k$ and $K_E^k$ being, respectively, the z-component of angular momentum and Carter's constant 
of motion, divided by energy.

Now we need to calculate the constants of motion $K_E^k$ and $L_E^k$ for a spherical light wave at $r_P$.  Using  \eqref{eq:Kiselev-L_E}  we obtain 
\begin{equation}
K_E^k(r_p)=\frac{16 r_p^2 \left(a^2+r_p \left(-k r_p^{-3 \alpha }-2 m+r_p\right)\right)}{\left((3 \alpha -1) k r_p^{-3 \alpha }-2 m+2 r_p\right)^2} ,
    \label{eq:Kiselev-K_E-final}
\end{equation}
\begin{equation}
\begin{aligned}
L_E^k(r_p)=&\resizebox{!}{20pt}{[}a^2 \left(-3 \alpha  k+k+2 (m+r_p) r_p^{3 \alpha }\right) +r_p^2 \left(2 (r_p-3 m) r_p^{3 \alpha }-3 (\alpha +1) k\right)\resizebox{!}{20pt}{]}\\
& \times \left({a (1-3 \alpha ) k+2 a (m-r_p) r_p^{3 \alpha }}\right)^{-1} . 
\end{aligned}
    \label{eq:Kiselev-L_E-final}
\end{equation}
Let us now introduce
\begin{equation}
    \Delta_k(r_0) =a^2+ k   r_0^{1-3 w}-2 m  r_0+r_0^2 ,
    \label{eq:delta-observer} 
\end{equation}
where ``$r_0$" denotes the position of astrophysical observer with respect to black hole. From \eqref{eq:sin-psi-3} and \eqref{eq:sin-zeta} we have
\begin{eqnarray}
     &&\psi_k ({r_p})=\sin ^{-1}\left(\frac{L_E^k(r_p)-a  \sin ^2(\theta_0 )}{\sqrt{K_E^k(r_p)} \sin ^2(\theta_0 )}\right) ,
    \label{eq:Kiselev-psi}
    \\ \nonumber\\
    &&\zeta_k ({r_p})=\sin^{-1}\left(\frac{\sqrt{\Delta_k(r_0)    K_E^k(r_p)}}{a^2-a   L_E^k(r_p)+r_p^2}\right) .
    \label{eq:Kiselev-zeta}
\end{eqnarray}


\begin{figure*}[t]
    \centering
      
    \begin{minipage}{0.26\textwidth}
        \centering
        \includegraphics[width=\linewidth]{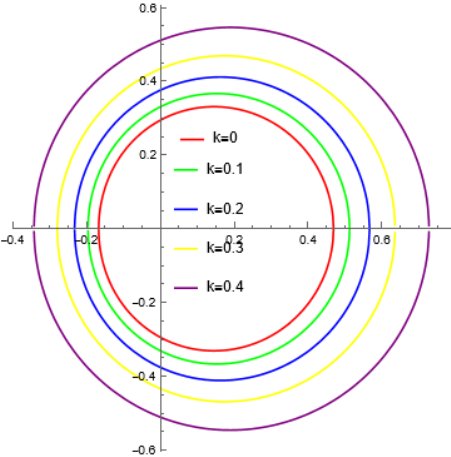}
        \caption*{(a) Dark Matter}
    \end{minipage}
    \hspace{15pt}
    \begin{minipage}{0.26\textwidth}
        \centering
        \includegraphics[width=\linewidth]{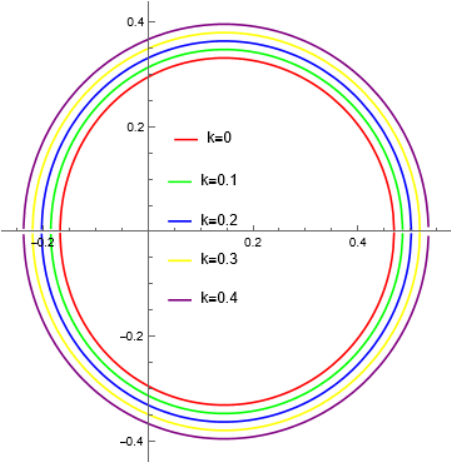}
        \caption*{(b) Dust}
    \end{minipage}
    \hspace{15pt}
    \begin{minipage}{0.26\textwidth}
        \centering
        \includegraphics[width=\linewidth]{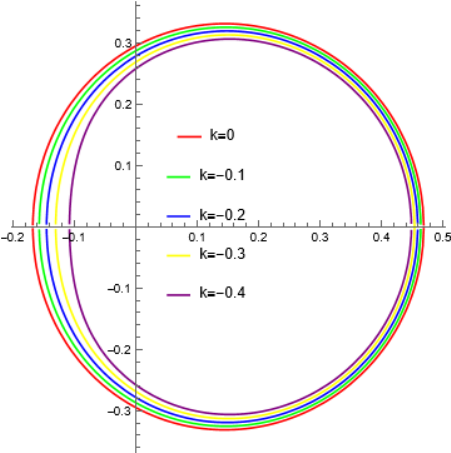}
        \caption*{(c) Radiation}
    \end{minipage}

    \caption{Shadows for different values of $\alpha$: (a) Dark Matter with $\alpha=-\frac{1}{3}$, (b) Dust with $\alpha=0$, and (c) Radiation with $\alpha=\frac{1}{3}$. The spin value is set to $a= 0.75m$ for both rotating Kiselev metrics and the standard Kerr spacetime. The observer is at $r_0= 15 m$ in the equatorial plane $\theta_0=\frac{\pi}{2}$ (we set $m=1$).}
    \label{fig:differentk-values}

\end{figure*}

Finally, we use the stereographic projection in \eqref{eq:plot-shadow} that maps the celestial sphere of the observer into Cartesian coordinates. 
Using \eqref{eq:Kiselev-psi} and \eqref{eq:Kiselev-zeta} we have

\begin{equation}
    X_k({r_p})\text{=}-2 \tan \left(\frac{\zeta_k (r_p)}{2}\right) \sin (\psi_k (r_p)) ,
\qquad
    Y_k({r_p})\text{=}-2 \tan \left(\frac{\zeta_k (r_p)}{2}\right) \cos (\psi_k (r_p)) .
\label{eq:Y_k}
\end{equation}

We numerically evaluate   $X_k({r_p})$ and $Y_k({r_p})$ and compare the resulting shadows with the shadow of the standard Kerr black hole. $r_p$ ranges from the photon sphere outward, as photons with smaller $r_p$ fall into the black hole, while those with larger $r_p$ define the shadow boundary.

The \figurename{ \ref{fig:Dark-Matter-plots-2}}, represents the case of a black hole surrounded by \textbf{dark matter}, corresponding to the choice of $\alpha=-\frac{1}{3}$ with $k>0$. It shows a significant deviation in comparison to the standard Kerr metric for different spin values. The cases with $k>0$ result in bigger shadow sizes compared to the standard Kerr metric, and the most pronounced deformation is for $a=0.85$

The \figurename{ \ref{fig:Dust-plots-2}}, represents the case of a black hole surrounded by \textbf{dust} corresponding to the choice of $\alpha=0$, with $k>0$. 
We recall that this case is equivalent to Kerr with
a shifted mass, $2m\to 2m+k$. Nevertheless, it is very useful to exhibit the $k$-dependence 
 for comparison with the other cases.

The \figurename{ \ref{fig:rad-plots-1}}, illustrates the case of a black hole surrounded by \textbf{radiation} corresponding to the choice of $\alpha=+\frac{1}{3}$ and $k<0$ (as shown in the previous section, in the radiation  case, $k<0$
is required by energy positivity). The figures show that, for different spin values, the deviation is less significant in comparison to the other cases. Note that  Finally, it is evident that cases with $k<0$ result in shadows of a smaller size than those of the Kerr, although the deformations are challenging to distinguish.

In \figurename{ \ref{fig:differentk-values}}, we tried to compare the effect of the parameter $k$, which represents the density of fluid matter enclosing the black hole. The effect of varying the parameter $k$ is examined, with the results compared to those obtained in the standard Kerr case. Note, here $k\equiv \kappa$ for convenience.

The figures presented in \figurename{ \ref{fig:differentk-values}}, illustrate the impact of varying fluid matter \textbf{density parameter $k$} on the shadow where $k=0$ refers to standard Kerr. The case of radiation demonstrates the minimal differences observed for varying values of the density parameter, $k$. When $a=0.75 m$ is fixed, the resulting shadows are nearly indistinguishable from a standard Kerr metric. In contrast, the black holes enclosed by dark matter exhibits a high degree of sensitivity to changes in the $k$ parameter. In the case of dust, the sensitivity is less apparent for different values of $k$ in comparison to the case of dark matter, which shows a high level of deformation especially for $k=0.4$.

 Upon examination of the right-hand side of the shadow curves, it becomes evident that the dark matter
 case exhibits a markedly increased sensitivity to alterations in the value of $k$, whereas the dust case demonstrates a barely discernible deformation of the curve for 
 each value of $k$.

\bigskip

\subsection{Bardeen's Distant Observer Method : Kiselev Metric}

Bardeen's method provides a convenient parametrization for distant observers as it uses impact parameters $(\vartheta_k , \beta_k)$ defined at infinity, directly mapping photon trajectories onto the observer’s sky, therefore closely resembling actual astronomical observations.

The Bardeen's dimensionful variables  $(\vartheta_k, \beta_k)$ are defined in terms of 
the dimensionless variables $(X_k, Y_k)$ in \eqref{eq:X_k}, \eqref{eq:Y_k}, as 
\begin{equation}
\vartheta_k=r_0 X_k(r_p) - a \sin \theta_0, \quad \beta_k=r_0 Y_k(r_p) .
\label{eq:dimnesionful-variables-bardeen-kiselev}
\end{equation}
Using \eqref{eq:x-plot-distant-observer}, \eqref{eq:y-plot-distant-observer}, we find
\begin{equation}
\begin{aligned}
	 & \vartheta_k\left(r_p\right) = -\frac{L^k_E(r_p)}{\sin \theta_0} ,\\
\end{aligned}
\label{eq:alpha-1-Kiselev}
\end{equation}
\begin{equation}
\begin{aligned} 
	& \beta_k (r_p)= \pm \left(K^k_E(r_p)-\left(L^k_E (r_p)-a\right)^2\right.\left.+a^2 \cos ^2 \theta_0-{L^k_E}^2 \cot^2 \theta_0\right)^{1 / 2} .
\end{aligned}
\label{eq:beta-1-Kiselev}
\end{equation}

\noindent By introducing new constants 
$$
\xi_k(r_p) \equiv \frac{L_k(r_p)}{E}\, ,\ \ \eta_k(r_p) \equiv K^k_E(r_p)-\left(L^k_E(r_p)-a\right)^2 ,
$$

\begin{figure*}[t]
    \begin{minipage}{0.26\textwidth}
        \centering
        \includegraphics[width=\linewidth]{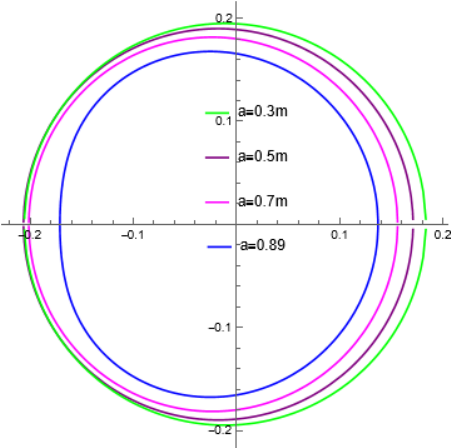}
        \caption*{(a) Dark Matter}
    \end{minipage}
    \hspace{15pt}
    \begin{minipage}{0.26\textwidth}
        \centering
        \includegraphics[width=\linewidth]{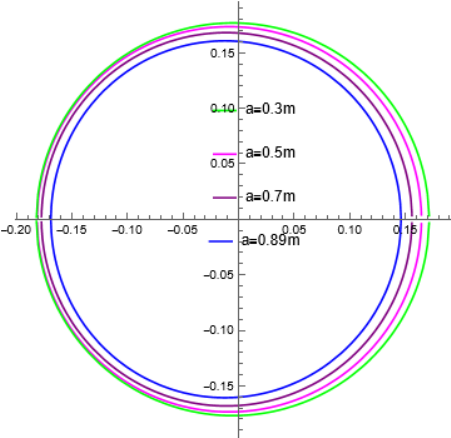}
        \caption*{(b) Radiation}
    \end{minipage}
    \hspace{15pt}
    \begin{minipage}{0.26\textwidth}
        \centering
        \includegraphics[width=\linewidth]{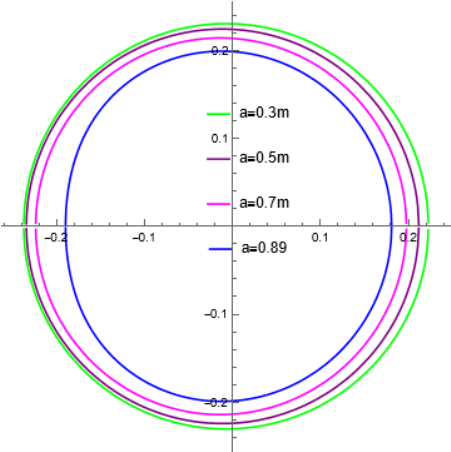}
        \caption*{(c) Dust}
    \end{minipage}

    \caption{Shadows using Bardeen's Method: (a)  $\alpha=-\frac{1}{3}$, $\kappa=0.2$ (b)  $\alpha=\frac{1}{3}$, $\kappa=-0.2$ and (c) $\alpha=0$, $\kappa=0.2$ (in all cases $m=1$, $\theta_0=\frac{\pi}{2}$).}
    
    \label{fig:Bardeen-Kiselev}

\end{figure*}

the equation \eqref{eq:beta-1-Kiselev} can be expressed as
\begin{equation}
\begin{aligned}
	 \beta_k(r_p)= \pm \eta_k\left(r_p\right)+a^2 \cos ^2 \theta_0-\xi_k^2\left(r_p\right) \cot^2\left(\theta_0\right) .
\end{aligned}
\label{eq:beta-final-Kiselev}
\end{equation}
Using the equation \eqref{eq:alpha-1-Kiselev} and \eqref{eq:beta-final-Kiselev} one can write the shadow curve as an explicit function $\beta(\vartheta)$,
\begin{equation}
    \begin{aligned}
        &\beta_k (\vartheta_k)  =\pm \sqrt{\eta_k(r_p)+ a^2 \cos ^2 \theta_0-\vartheta_k^2 \sin ^2 \theta_0\left(\frac{\cos ^2 \theta_0}{\sin ^2 \theta_0}\right)} .
    \end{aligned}
    \label{eq:beta-terms-alpha-Kiselev}
\end{equation}
In order to express this function in terms of constants of motion, we first
substitute  \eqref{eq:Kiselev-K_E-final} and  \eqref{eq:Kiselev-L_E-final} into  $\eta_k(r_p)$, $\xi_k(r_p)$. This gives
\begin{equation}
\begin{aligned}
    \setlength{\belowdisplayskip}{10pt}
  \xi_k(r_p) =  &\resizebox{!}{20pt}{(} k \left((3 \alpha -1) a^2+3 (\alpha +1) r_p^2\right) +2 r_p^{3 \alpha } \left(a^2 (m+r_p)+r_p^2 (r_p-3 m)\right)\resizebox{!}{20pt}{)}\\
  &\times \left[a (3 \alpha -1) k+2 a (m-r_p) r_p^{3 \alpha }\right]^{-1} ,
 \end{aligned}
  \label{lambda-kiselev}
\end{equation}
\begin{equation}
\begin{aligned}
    \eta_k(r_p) =-&\resizebox{!}{18pt}{(} 4 k r_p^{3 \alpha +3} \left(a^2 (6 \alpha +5)-3 (\alpha +1) r_p (3 m-r_p)\right) +4 r_p^{6 \alpha +2} \left(3 a^4+a^2 r_p (3 r_p-10 m)+r_p^2 (r_p-3 m)^2\right) \\
    &+ 9 (\alpha +1)^2 k^2 r_p^4\resizebox{!}{18pt}{)}  \left(a^2 \left(-3 \alpha  k+k-2 (m-r_p) r_p^{3 \alpha }\right)^2\right)^{-1}. 
\end{aligned}
\label{eta_kiselev}
\end{equation}
We numerically solve \eqref{lambda-kiselev} for $r_p$ and substitute it into
\eqref{eta_kiselev}, obtaining $\eta_k(\xi_k)$. 
This is then used to plot $\beta_k(\vartheta_k)$ for the different cases of fluid matters: \text{Dark Matter}, \text{Dust} and \text{Radiation} corresponding to equations of states $\alpha=-\frac{1}{3}$, $\alpha=0$ and $\alpha=\frac{1}{3}$ respectively.  

The results are shown in \figurename{ \ref{fig:Bardeen-Kiselev}} and they confirm the pattern observed in the shadows 
of previous figures. The shape in the dark matter case is clearly distinguishable from the other two cases, especially for large values of the rotation parameter $a$.

The deformations of the shadow due to the presence of Dark matter are more visible, even for relatively low spin values compared to the dust case. On the other hand,  the case of radiation exhibits a behavior which is similar to the Kerr
case.  It is interesting to note that for $a=0.89m$ the shadow in the dark matter case is much smaller than that of
the radiation and dust cases. This distinction serves to differentiate the dark matter case from the other cases under consideration.

 \section{QNMs of Kiselev Black Hole}\label{ch:QNM}

The metric perturbations include modes with spin 0, 1 and spin 2. The spin 0 perturbations represent the scalar metric perturbations and are governed by the 
wave equation of a scalar field \cite{Dolan_2022}.
In this section we will focus on scalar metric perturbations \cite{Khodadi_2021}.
	The wave equation for a massless scalar field \(\Phi\)  is given by
	\begin{equation}
 \begin{aligned}
	 \left(\frac{1}{\sqrt{-g}} \partial_\alpha\left(\sqrt{-g} g^{\alpha \beta} \partial_\beta\right)\right) \Phi(t, r, \theta, \phi) = 0\, .
  \end{aligned}
		\label{eq:kg_equation}
	\end{equation}
We consider the following ansatz in the Boyer-Lindquist coordinates \((t, r, \theta, \phi)\), 
	\begin{align}
		&\Phi(t, r, \theta, \phi) = R_{\omega l m_l}(r) S_{\omega l m_l}(\theta) e^{-i \omega t} e^{i m_l \phi}\, , \qquad l \geq 0\ ,\quad -l \leq m_l \leq l, \quad \omega>0\, . 
		\label{eq:ansatz}
	\end{align}
	Here \(l\), \(m_l\), and \(\omega\) denote the angular quantum number, azimuthal wave number and positive frequency of the scattering scalar field observed by a distant observer. Substituting the metric \eqref{eq:Kiselev-Metric-QNM} into the differential equation \eqref{eq:kg_equation} and using the ansatz \eqref{eq:ansatz}, we obtain two  differential equations
    for the radial and angular parts:
	\begin{align}
		& \frac{d}{d r}\left(\Delta_k \frac{d R_{\omega l m_l}(r)}{d r}\right)  + \left(\frac{\left(\left(r^2+a^2\right) \omega-a m_l\right)^2}{\Delta_k}\right. -\left(l(l+1)+a^2 \omega^2-2 m_l a \omega\right)\resizebox{!}{22pt}{)} R_{\omega l m_l}(r)=0 ,
		\label{eq:radial_eqn}
	\end{align}
	\begin{align}
		& \sin \theta \frac{d}{d \theta}\left(\sin \theta \frac{d S_{\omega l m_l}(\theta)}{d \theta}\right) + \resizebox{!}{20pt}{(}l(l+1) \sin ^2 \theta-\left(\left(a \omega \sin ^2 \theta-m_l\right)^2\right)\resizebox{!}{20pt}{)} S_{\omega l m_l}(\theta)=0 .
  \label{eq:angular_eqn}
	\end{align}
    By introducing a ``tortoise" coordinate \(r_*\) as \(\frac{d r_*}{d r} \equiv \frac{r^2+a^2}{\Delta_k}\) (with \(r_* \rightarrow-\infty\) at the event horizon and \(r_* \rightarrow \infty\) at infinity) and defining a new radial function \(\mathcal{F}_{\omega l m_l}\left(r_*\right)=\sqrt{r^2+a^2} R_{\omega l m_l}(r)\), the radial equation \eqref{eq:radial_eqn} becomes  a Schrödinger-like differential equation:
	\begin{equation}
		\frac{d^2 \mathcal{F}_{\omega l m_l}\left(r_*\right)}{d r_*^2}+U_{\omega l m_l}(r) \mathcal{F}_{\omega l m_l}\left(r_*\right)=0 ,
		\label{eq:schrodinger_eqn}
	\end{equation}
	where
	\begin{align}
		&U_{\omega l m_l}(r) = (\omega-m_l \Omega)^2-\frac{\Delta_k\left(l(l+1)+a^2 \omega^2-2 m_l a \omega\right)}{\left(r^2+a^2\right)^2}-\frac{\Delta_k\left(3 r^2-4 m r+a^2\right)}{\left(r^2+a^2\right)^3}+\frac{3 r^2 \Delta_k }{\left(r^2+a^2\right)^4} ,
		\label{eq:potential}
		\end{align}
		is the scattering potential. Here, \(\Omega=\frac{a}{r^2+a^2}\) represents the angular velocity of the rotating BH. The coordinate $r_*$ maps the range
  $r \in [ r_+, \infty) $ to the whole real axis.
  
  The boundary conditions on the event horizon and
  at infinity are as follows
\begin{equation}
R_{\omega l m_l}(r) \rightarrow \begin{cases}\mathcal{I}^{+} \frac{e^{-i k_{+} r_*}}{\sqrt{r^2+a^2}} & \text { for } r \rightarrow r_{+}\left(r_* \rightarrow-\infty\right) \\ \mathcal{I}^{\infty} \frac{e^{-i k_{\infty} r_*}}{r} & \text { for } r \rightarrow \infty \quad\left(r_* \rightarrow \infty\right)\end{cases} ,
\label{eq:boundary-condition-R}
\end{equation}
 while 
  \begin{equation}
\begin{aligned}
& \lim _{r \rightarrow r_{+}} U_{\omega l m_l}(r)=\left(\omega-m_l \Omega_{+}\right)^2, \quad \Omega_{+}\equiv\frac{a}{r_{+}^2+a^2}; \\
& \lim _{r \rightarrow \infty} U_{\omega l m_l}(r)=\omega^2\  .
\end{aligned}
\label{eq:boundary-conditions-U}
\end{equation}

\subsection{Eikonal Approximation}

The eikonal approximation involves taking a large angular momentum limit, $\ell \gg 1$.  This approximation simplifies the wave equations significantly.  A similar approach was  used in \cite{Dolan} for Kerr black holes.


The QNMs are required to satisfy the following boundary conditions 
\begin{equation}
\mathcal{F}_{\omega l m_l}\left(r_*\right) = C_{\pm} \exp\left(\pm i \omega r_{\star}\right), \quad r_* \rightarrow \pm \infty, \label{eq:boundary_condition}
\end{equation}
where $\omega$ can be expressed as a combination of its real and imaginary parts, $\omega = \omega_{\Re} + i \omega_{\Im}$, with $\omega_{\Re}$ representing oscillation frequency, and $\omega_{\Im}$ being proportional to the decay rate of the mode.

In the eikonal limit, the potential \eqref{eq:potential} becomes
\begin{equation}
   \lim_{l\rightarrow \infty} U_{\omega l m_l}(r) \equiv \mathcal{U} = - \left(\frac{a^2-k r^{1-3 \alpha }-2 m r+r^2}{\left(a^2+r^2\right)^2} \right) l^2 
   \label{eq:limit-potential-u}
\end{equation}
which is independent of $m_l$. 
The extremum of $\mathcal{U}$ at \(r_0\) can be determined
by solving $\frac{d \mathcal{U}}{d r_*}= 0$ . 
We find
	\begin{align}
		&\frac{d \mathcal{U}}{d r_*} =
		\resizebox{!}{20pt}{(}\frac{4 r \left(a^2-k r^{1-3 \alpha }-2 m r+r^2\right)}{\left(a^2+r^2\right)^3}-\frac{(1-3 \alpha ) (-k) r^{-3 \alpha }-2 m+2 r}{\left(a^2+r^2\right)^2}\resizebox{!}{20pt}{)} l^2  .
		\label{eq:dU_dr}
	\end{align}
 This leads to real solutions  \(r_0^{\text{Rad}}\), \(r_0^{\text{DM}}\), and \(r_0^{\text{Dust}}\) corresponding to the cases $\alpha = \frac{1}{3}$, $\alpha = -\frac{1}{3}$ and $\alpha = 0$ respectively.  
 
 Let us define $K_0$ as
\begin{equation}
\tilde{\mathcal{U}}_0\equiv \frac{\mathcal{U}}{l^2}\bigg|_{r=r_0}\ .
\end{equation}



\noindent In the eikonal approximation the QNM frequencies depend on quantum numbers $l$ and $n$ -- the principal quantum number or overtone number. In addition, frequencies depend on BH parameters  $a$, $k$, $\alpha$.
We will focus on modes zero overtone number, as these  have significantly longer  lifetimes.
The frequency for modes with $n=0$ can be expressed analytically as \cite{Glampedakis}:
	\begin{align}
		\omega_{\Re} & = l \sqrt{\tilde {\mathcal{U}}_0}+O\left(l^0\right),  \\
		\omega_{\Im} & = -\frac{1}{2}\left(\frac{d r}{d r_*}\right)_0 \sqrt{\frac{\left|{\tilde{\mathcal{U}}}_0^{(2)}\right|}{2 {\tilde{\mathcal{U}}}_0}}+O\left(l^{-1}\right).
		\label{eq:frequency}
	\end{align}
Thus, for $l\gg 1$, the real part of the frequency grows linearly with the angular momentum and the imaginary part approaches a constant value.

 \begin{figure*}
    \centering
    
    \begin{subfigure}
        \centering
        \begin{minipage}{0.45\textwidth}
            \centering
            \includegraphics[width=\linewidth]{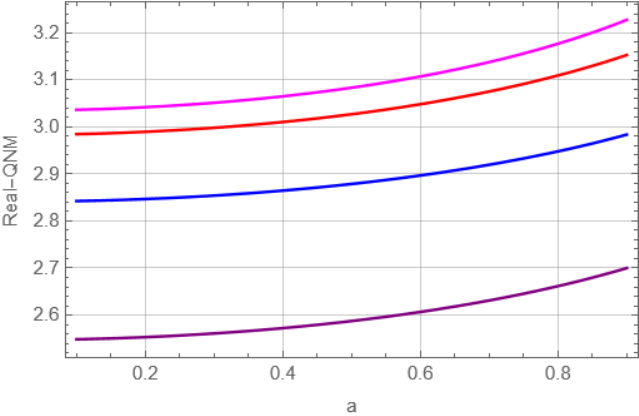}
            \caption{$\omega_{\Re}$ vs. $a$.  
            \textcolor{red}{Kerr} - \textcolor{magenta}{Radiation $\alpha=\frac{1}{3}$} - \textcolor{blue}{Dust $\alpha=0$} - \textcolor{dark purple}{Dark Matter $\alpha=-\frac{1}{3}$}. We fixed the parameters $l = 15$ and $|\kappa| = 0.1$.}
            \label{fig:Real-QNM-a}
        \end{minipage}
        \hspace{5pt}
        \begin{minipage}{0.45\textwidth}
            \centering
            \includegraphics[width=\linewidth]{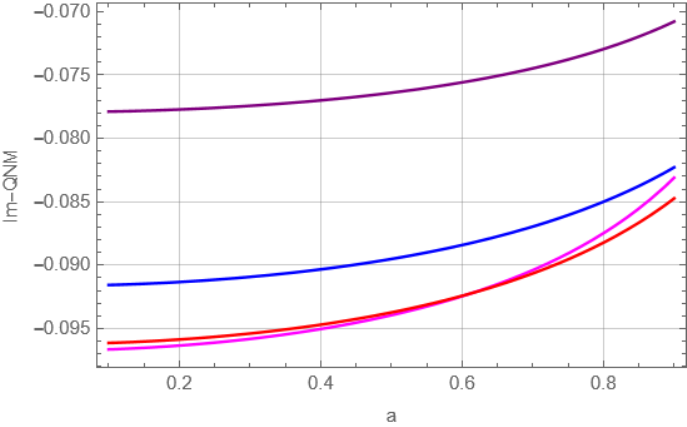}
            \caption{$\omega_{\Im}$ vs. $a$. \textcolor{red} {Kerr} - \textcolor{magenta}{Radiation $\alpha=\frac{1}{3}$} - \textcolor{blue}{Dust $\alpha=0$} - \textcolor{dark purple}{Dark Matter $\alpha=-\frac{1}{3}$}. We fixed the parameters $l = 15$ and $|\kappa| = 0.1$.}
            \label{fig:Img-QNM-a}
        \end{minipage}
    \end{subfigure}
    \vspace{15pt}
    \begin{subfigure}{ }
        \centering
        \begin{minipage}{0.45\textwidth}
            \centering
            \includegraphics[width=\linewidth]{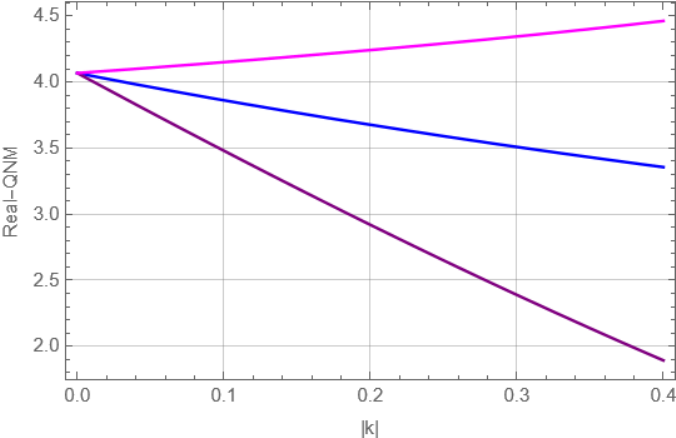}
            \caption{$\omega_{\Re}$ vs. $|k|$. \textcolor{magenta}{Radiation $\alpha=\frac{1}{3}$} - \textcolor{blue}{Dust $\alpha=0$} - \textcolor{dark purple}{Dark Matter $\alpha=-\frac{1}{3}$}. We fixed the parameter $l = 20$, $a=0.7$ and $m=1$.}
            \label{fig:real-QNM-k-p}
        \end{minipage}
        \hspace{5pt}
        \begin{minipage}{0.45\textwidth}
            \centering
            \includegraphics[width=\linewidth]{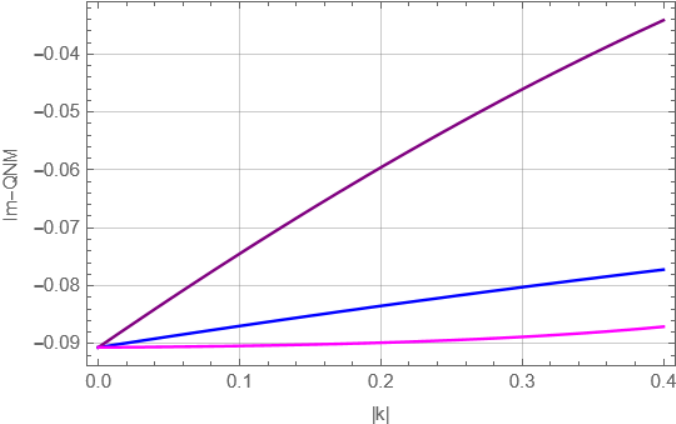}
            \caption{$\omega_{\Im}$ vs. $|k|$. \textcolor{magenta}{Radiation $\alpha=\frac{1}{3}$} - \textcolor{blue}{Dust $\alpha=0$} - \textcolor{dark purple}{Dark Matter $\alpha=-\frac{1}{3}$}. We fixed the parameter $l = 20$, $a=0.7$ and $m=1$.}
            \label{fig:img-QNM-k-p}
        \end{minipage}
    \end{subfigure}
    \vspace{10pt}
\end{figure*}

The dependence of $\omega$ on the spin parameter $a$ and on the density parameter $|k|$ exhibits some interesting features.
The real and imaginary parts of the frequencies  are plotted in terms of the spin parameter $a$ in  \figurename{ \ref{fig:Real-QNM-a} and  \figurename{ \ref{fig:Img-QNM-a}} for $l=15$, $n=0$, setting \(  m=1\) and $\lvert  \kappa \rvert  = 0.1$, where $k=\kappa m^{1+3\alpha}$ ($\kappa$ is negative for $\alpha=\frac{1}{3}$, and positive for $\alpha=0$ and $\alpha=-\frac{1}{3}$ ). The figures 
compare the cases of radiation, dust and dark matter with the frequencies for $k=0$ corresponding to the Kerr black hole.
The largest deviation occurs for $\alpha=-1/3$.
Notably, the  $\alpha=1/3$ case exhibits values that are very close  to the Kerr case. 

The real part represents the actual frequency of  oscillation for gravitational waves. 
This suggests that for a Kiselev geometry with $\alpha=-1/3$  one might expect an enhancement of the signal, since gravitational waves at lower frequencies are easier to detect.
The \figurename{ \ref{fig:Img-QNM-a}}, illustrates that the cases of Kerr and radiation exhibit a remarkably similar behavior. It is important to note that the imaginary parts of the frequencies are negative  ($\omega_{\Im}<0$), corresponding
to decay modes with lifetimes proportional to $|\omega_{\Im}|^{-1}$.  The case of dark matter exhibits values of $\omega_{\Im}$ that are notably closer to zero. This suggests that in the case of dark matter  may be  more astrophysically viable for detection purposes, since the lifetime of the modes  will be greater than that of the other cases. 
 \figurename{ \ref{fig:Img-QNM-a}} also shows that
 in all  cases the lifetimes of the QNMs increase
 as the spin parameter $a$ is increased.

 It can thus be concluded that quasi-normal modes associated with near-extremal black holes have greater potential for precision astrophysical experiments.

  \figurename{ \ref{fig:real-QNM-k-p}} and \figurename{ \ref{fig:img-QNM-k-p}} show the behavior of the real and imaginary parts of the QNM frequecies as a function of the density parameter $k$.
  The real part of the frequency exhibits a clearly distinguishable behavior for different cases of fluid matter. 
  In the case of radiation, the real part of the frequency increases as $|k|$ increases, whereas the opposite behavior is
  seen for the case of Dust and Dark matter. The observed discrepancy in behavior is presumably attributable to the fact that an increase in $|k|$ for radiation corresponds to a shift towards extremality, while an increase in $|k|$ for dust is associated with a departure from extremality.
  The figure also shows that the real part of the frequency
  is more sensitive to the change in $k$ in the dark matter case, since
  the slope is significantly steeper. As a result, in the dark matter case, the real part of the frequency decreases more rapidly as the density of fluid matter increases.
This may also have an implication for astrophysical observations, since 
 gravitational waves with lower frequencies can be more easily identified from the signals.
 
 \figurename{ \ref{fig:img-QNM-k-p}} shows that the lifetime of QNMs (proportional to $|\omega_\Im|^{-1}$) increases with $|k|$ in all cases, but
 with a notably higher increase in the case of Dark matter.
 Again, this feature may be of relevance for astrophysical observations since  longer lifetimes facilitates the analysis of QNMs.

\section{Conclusion} 

In this paper we have studied the effect of deformations of the Kerr metric on black hole shadows
and on quasinormal modes. The two-parameter family of deformations correspond to a solution of Einstein equations known as the Kiselev rotating black hole,
where $k$ is a parameter controlling the energy density and $\alpha $ controlling the asymptotic behavior of the stress-energy tensor.
As prototypical examples, we focused on three cases, $\alpha=\{ -\frac13, \, 0,\ \frac13\}$, which mimic
the  stress-energy tensor for certain models of dark matter, dust and radiation, respectively.

The black hole shadows have been determined in a wide range of parameters.
In  \figurename{ \ref{fig:Dark-Matter-plots-2}} to \figurename{ \ref{fig:rad-plots-1}}, we have shown
how the shadow behaves as the rotation parameter $a$ is increased, for the three cases of fluid matter, DM, radiation and dust. The figures show more significant deviations with respect to the Kerr black hole shadow
in the DM case. 

In \figurename{\ref{fig:differentk-values}} we study the deformation of the shadows for the different cases of fluid matter as the density parameter $k$ is gradually increased. Once again, the dark matter case shows greater sensitivity to the increase in the density parameter. This is in contrast to the radiation case, which shows little sensitivity despite the fact that the metric for $\alpha=\frac13$ approaches extremality for larger $|k|$.

 In \figurename{\ref{fig:Bardeen-Kiselev}}, we show the black hole shadows parametrized using  the alternative parametrization provided by the Bardeen's method.
 The figures compare the shadows for different values of the rotation parameter in the same graph. In all cases, increasing
 the rotation parameter reduces the size of the black hole shadow. Concurrently, for a rotation parameter of greater magnitude, a more pronounced deviation from a circular shape is exhibited by the shadows.
 This phenomenon is more pronounced in the cases of dark matter and dust, while it is almost imperceptible in the case of radiation.

The analysis of quasi-normal modes  for different types of fluid matter, across a range of parameters, unveils the following behavior. In all cases, the QNMs become more stable for higher density and for higher spin parameter $a$.
Furthermore, it is observed that the lifetimes of QNMs are significantly greater in the case of dark matter.

The real part of the frequency, controlling the oscillating part of QNMs, increases as the spin parameter increases, and it is notably smaller in the case of Dark Matter.
For fixed spin parameter, in the  DM and dust case, the real part of the frequency decreases as the energy density is increased,
while it increases in the case of radiation. We attributed this different behavior to the fact that in the case of radiation
the BH approaches extremality as $|k|$ increases (while
for dust and DM the BH departs from extremality as $k$ increases).

For potential astrophysical applications, these results suggest that a dominance of dark matter in the stress tensor surrounding black hole mergers could  favor gravitational wave detection, since low QNM frequencies are generally easier to detect with current gravitational wave detectors. The design and sensitivity of these detectors are optimized for lower frequency ranges, where astrophysical signals are stronger and the noise is lower, making it more likely to distinguish the characteristic oscillations due to QNMs.
In the case of dark matter, QNMs  have, in addition, a longer lifetime,  making this case more favorable for detection purposes.

It is conceivable that forthcoming gravitational wave experiments or black hole imaging may yield outcomes that deviate from the predictions derived from Kerr black hole geometries. The present results provide a precise account on how the behavior of quasinormal modes and black hole shadows is affected by the presence of stress-energy tensors with specific asymptotic behaviors at infinity. 
This complements other related results in the literature for other classes of black hole deformations.
It is clear that a more extended analysis of black hole deformations will be required to understand the underlying origin of any potential deviations from the predictions based on Kerr black hole geometries.

\section*{Acknowledgements}

RP thanks the Institute of Cosmos Sciences at the University of Barcelona for hospitality during the course of this work.
JGR acknowledges financial support from grant 2021-SGR-249 (Generalitat de Catalunya) and  by the Spanish  MCIN/AEI/10.13039/501100011033 grant PID2022-126224NB-C21.

\small
 \bibliographystyle{apsrev4-2}
\bibliography{references}
\clearpage

\begin{widetext}

    \appendix

\section{Shadow of Rotating Kiselev Black Holes}\label{sec:rotating-BH}

In this appendix we provide a detailed derivation
of the shadow boundary curve for the rotating Kiselev black hole.
We shall follow the same notation as in  \cite{PERLICK2022}. In order to study geodesic motion, we start with  the Lagrangian for the   rotating Kiselev geometry,
\begin{equation}
\begin{aligned}
	 & \mathcal{L}=\frac{1}{2} g_{\mu \nu} \dot{x}^\mu \dot{x}^\nu =\frac{1}{2}\left[-\frac{\Delta_k-a^2 \sin ^2 \theta}{\Sigma} \dot{t}^2+\frac{\Sigma}{\Delta_k} \dot{r}^2+\Sigma \dot{\theta}^2\right. \\
	& \left.+\frac{\sin ^2 \theta}{\Sigma}\left(\left(r^2+a^2\right)-\Delta_k a^2 \sin ^2 \theta\right) \dot{\phi}^2+\frac{2 a \sin ^2 \theta}{\Sigma}\left(\left(r^2+a^2\right)-\Delta_k\right) \dot{\phi} \dot{t}\right],
\end{aligned}
\label{eq:lag-Kerr}
\end{equation}
where $\Delta_k = r^2 - 2 m r +a^2 - k r^{1-3 \alpha} $ and $\Sigma=r^2+a^2 \cos ^2 \theta$. 
In the null case we have $\mathcal{L}=0$.
 The energy is given by
\begin{equation}
E=-\frac{\partial \mathcal{L}}{\partial \dot{t}}=\frac{\Delta_k-a^2 \sin ^2 \theta}{\Sigma} \dot{t}-\frac{a \sin ^2 \theta}{\Sigma}\left(\left(r^2+a^2\right)-\Delta_k\right) \dot{\phi},
\label{eq:E-Kerr}
\end{equation}
and the $z$-component of the angular momentum is
\begin{equation}
L_z=\frac{\partial \mathcal{L}}{\partial \dot{\phi}}=\frac{\sin ^2 \theta}{\Sigma}\left(\left(r^2+a^2\right)-\Delta_k a^2 \sin ^2 \theta\right) \dot{\phi}+\frac{a \sin ^2 \theta}{\Sigma}\left(\left(r^2+a^2\right)-\Delta_k\right) \dot{t}.
\label{eq:Lz-Kerr}
\end{equation}

\subsection{Carter's Constant}

In order to rewrite all four geodesic equations as first order equations, we need another conserved quantity found in 1968 by Carter. 
It is convenient to use  the Hamilton-Jacobi formalism.
Recall that a sufficiently general solution $s\big(\lambda, x^\mu\big)$ to the Hamilton-Jacobi equation satisfies
\begin{equation}
\begin{gathered}
	\frac{\partial s}{\partial \lambda}+H(d s)=0 ,
\end{gathered}
\label{eq:Hamilton-Jacobi}
\end{equation}
which determines the momentum of trajectories via $p \equiv d s$. 
When the Hamilton -Jacobi equation can be separated, the separation constant represents constant of motion.
Using \eqref{eq:lag-Kerr}, the Hamilton-Jacobi equation 
can be expressed as
%
%
\begin{equation}
\begin{aligned}
	 & -\Sigma \frac{\partial s}{\partial \lambda}=-\frac{1}{2 \Delta_k}\left[\left(r^2+a^2\right) \frac{\partial s}{\partial t}+a \frac{\partial s}{\partial \phi}\right]^2 
	+  \frac{\Delta_k}{2}\left(\frac{\partial s}{\partial r}\right)^2+\frac{1}{2 \sin ^2 \theta}\left[\frac{\partial s}{\partial \phi}+a \sin ^2 \theta \frac{\partial s}{\partial t}\right]^2  +\frac{1}{2}\left(\frac{\partial s}{\partial \theta}\right)^2.
\end{aligned}
\label{eq:rearr-HJ-Carter}
\end{equation}

Now, taking into account the definitions of the constants of motion in \eqref{eq:E-Kerr}, \eqref{eq:Lz-Kerr}, the equation 
\eqref{eq:rearr-HJ-Carter} becomes
%
\begin{equation}
    \left.P_\theta^2-2 a^2 \cos ^2 \theta H+\frac{1}{\sin ^2 \theta}\left[L_{z}-a E \sin ^2 \theta\right]^2\right.  =\left.-\Delta_k P_r^2+2 H r^2+\frac{1}{\Delta_k}\left[L_{z} a-E\left(r^2+a^2\right)\right]^2\right. .
\label{eq:separ-Carter1}
\end{equation}
Equation \eqref{eq:separ-Carter1} is now separable. By introducing a separation parameter $C$,
we have for the angular and radial parts the following 
equations
\begin{equation}
\begin{aligned}
   &C=P_\theta^2-2 H {a^2 \cos ^2 \theta} + \frac{1}{\sin ^2 \theta}\left[L_{z}-a E \sin ^2 \theta\right]^2, \\
	&C  =-\Delta_k P_r^2+2 H r^2+\frac{1}{\Delta_k}\left[L_{z} a-E\left(r^2+a^2\right)\right]^2 .
\end{aligned}
\label{eq:separ-radial}
\end{equation}
Since functions of conserved quantities are obviously conserved, any function of $C$ and the  other constants of the motion can be used as a new constant in place of $C$.
It is   convenient to define
\begin{equation}
	K=C - \left(L_z-a E\right)^2,
 \label{eq:K-Carter}
\end{equation}

in place of $C$, where $K$ denotes Carter's constant. 
The quantity $K$ is useful because it is always non-negative. 
Now, using \eqref{eq:K-Carter} into \eqref{eq:separ-radial}, one can rewrite the  equation \eqref{eq:separ-radial} for the angular variables as
    
\begin{equation}
\begin{aligned}
    K=P_\theta^2+\cos ^2 \theta\left[\frac{L_{z}^2}{\sin ^2 \theta}-a^2\left(2 H+E^2\right)\right]. 
\end{aligned}  
\label{eq:K-Carter-angular}
\end{equation}

\subsection{Geodesic Equations in a Rotating metric}

By using the contra-variant form of the rotating metric and using
$	H=\frac{1}{2} g^{\mu \nu} P_\mu P_\nu $,  we may express the Hamiltonian as 

\begin{equation}
\begin{aligned}
	&  H=-\frac{\left[\left(r^2+a^2\right)^2-\Delta_k a^2 \sin ^2 \theta\right]}{2 \Sigma \Delta_k} P_t^2  +\frac{\Delta_k}{2 \Sigma} P_r^2+\frac{1}{2 \Sigma} P_\theta^2+\frac{\Delta_k-a^2 \sin ^2 \theta}{2 \Sigma \Delta_k \sin ^2 \theta} P_\phi^2  -\frac{2 a\left[\left(r^2+a^2\right)-\Delta_k\right]}{2 \Sigma \Delta_k} P_t P_\phi .
 \end{aligned}
 \label{eq:Hamiltoni-Geodesic-1}
\end{equation}
From $\frac{\partial H}{\partial P_i}=\dot{q}_i $ we have $\dot{t}=\frac{\partial H}{\partial P_t}$. Thus, we get
 \begin{equation}
\begin{aligned}
	& \dot{t}= -\frac{\left[\left(r^2+a^2\right)^2-\Delta_k a^2 \sin ^2 \theta\right]}{\Sigma \Delta_k} P_t-\frac{a\left[\left(r^2+a^2\right)-\Delta_k\right]}{\Sigma \Delta_k} P_\phi.
  \end{aligned}
 \label{eq:t-dot}
\end{equation}
Using the definitions of the constants of motion in \eqref{eq:E-Kerr} and \eqref{eq:Lz-Kerr}, this can be written as
\begin{equation}
\dot{t}=\frac{a \sin ^2 \theta\left(L_{z}-E a \sin ^2 \theta\right)}{\Sigma \sin ^2 \theta}+\frac{\left(\Sigma+a^2 \sin ^2 \theta\right)\left[\left(\Sigma+a^2 \sin ^2 \theta\right) E-a L_{z}\right]}{\Sigma \Delta_k} ,  
\label{eq:t-dot-2}
\end{equation}
%

Similarly, we now compute
$\dot{\phi}=\frac{\partial H}{\partial P_\phi}$. We get 
\begin{equation}
\begin{aligned}
	&  \dot{\phi}=\frac{\Delta_k-a^2 \sin ^2 \theta}{\Sigma \Delta_k \sin ^2 \theta} P_\phi-\frac{a\left[\left(r^2+a^2\right)-\Delta_k\right]}{\Sigma \Delta_k} P_t .
 \end{aligned}
 \label{eq:dot-phi-1}
\end{equation}
In terms of the constants of motion \eqref{eq:E-Kerr} and \eqref{eq:Lz-Kerr}, this becomes
\begin{equation}
\begin{aligned}
	 \dot{\phi}=\frac{L_{z}-E a \sin ^2 \theta}{\Sigma \sin ^2 \theta}+\frac{a\left[\left(\Sigma+a^2 \sin ^2 \theta\right) E-a L_{z}\right]}{\Sigma \Delta_k} .
\end{aligned}    
\label{eq:dot-phi-2}
\end{equation}

By introducing $x=a\sin^2 \theta$ the equations \eqref{eq:t-dot-2} and \eqref{eq:dot-phi-2} can be re-expressed as
\begin{equation}
     \begin{aligned}
		&\dot{t}=\frac{x(L_{z}-E x)}{\Sigma \sin ^2 \theta}+\frac{(\Sigma+a x)[(\Sigma+a x) E-a L_{z}]}{\Sigma \Delta_k} \\
		&\dot{\phi}=\frac{L_{z}-E x}{\Sigma \sin ^2 \theta}+\frac{a[(\Sigma+a x) E-a L_{z}]}{\Sigma \Delta_k}
	\end{aligned}
 \label{eq:dot-t-phi-interms-x}
\end{equation}
%
Defining $\Theta \equiv P_\theta^2=\Sigma^2 \dot{\theta}^2$, \eqref{eq:K-Carter-angular} takes the form
\begin{equation}
\begin{aligned}
	&  \Theta=K-\cos ^2 \theta\left[\frac{L_{z}^2}{\sin ^2 \theta}+a^2\left(-2 H-E^2\right)\right] .
\end{aligned} 
\label{eq:theta-K}
\end{equation}
 For $\dot{r}$, we can similarly 
 define $R\equiv \Sigma ^2\dot{r}^2$
 and
   \eqref{eq:separ-radial} becomes
 %
 \begin{equation}
\begin{aligned}
	 R=\left[E\left(r^2+a^2\right)-L a\right]^2-\Delta_k\left[(L-a E)^2-2 H r^2+K\right] .
\end{aligned}
\label{eq:radial-R}
\end{equation}
where we dropped subscript $z$ for $L_z$. These equations 
can be solved explicitly in terms of hyperelliptic functions. Here we are interested in spherical lightlike geodesics; e.g., lightlike geodesics that stay on a sphere $r=\text{const.}$ The region filled by these geodesics is called the Photon region. 

\medskip
 
\textbf{Lemma}

Null geodesics have at most one radial turning point outside the horizon.

\medskip

\textit{Proof :}

For null geodesics, $H=0$, and  \eqref{eq:radial-R} becomes

\begin{equation}
\begin{aligned}
	&  R(r)=\left[E\left(r^2+a^2\right)-a L\right]^2-\Delta_k\left[(L-a E)^2+K\right] .
 \end{aligned}
 \label{eq:R(r)-null-geodesics}
\end{equation}
Using $K=-\Delta_k P_r^2+\frac{1}{\Delta_k}\left[L a-E\left(r^2+a^2\right)\right]^2$ we get
\begin{equation}
\begin{aligned} R(r)=\Delta_k\left[\Delta_k P_r^2-(L-\alpha E)^2\right],
\end{aligned}    
\end{equation}
%
For Kiselev metric, the location of the horizon is at
 $\Delta_k=r^2-2 m r+a^2 - k r^{1 - 3 \alpha} =0$. This
 gives

    \begin{itemize}
    \item For $\alpha = \frac{1}{3}$:
    \[
    r^{\text{Rad-H}}_{\pm} = m \pm \sqrt{m^2-a^2 + k  }
    \]

    \item For $\alpha = -\frac{1}{3}$:
    \[
    r^{\text{DM-H}}_{\pm} = \frac{m \pm \sqrt{m^2-a^2 + a^2 k }}{1 - k}
    \]

    \item For $\alpha = 0$:
    \[
    r^{\text{Dust-H}}_{\pm} = \frac{ 2m +k\pm \sqrt{4 m^2-4 a^2 + k^2 + 4 k m }}{2}
    \]
\end{itemize}
The turning points occur at $\dot r=0$, i.e.  $R(r) = 0$.  We solve  for $\Delta_k \neq 0$ obtaining

\begin{equation}
\begin{aligned}
	&  \Delta_k =D\ ,\qquad 
  D\equiv \frac{(a E-L)^2}{P_r^2}\geq 0 \ .
\end{aligned}    
\end{equation}

Hence 
\begin{equation}
\begin{aligned}
    \text{For } \alpha = \frac{1}{3}, \quad r^{\text{Rad}}_{\pm} &= m \pm \sqrt{m^2-a^2 + D + k }, \\
    \text{For } \alpha = -\frac{1}{3}, \quad r^{\text{DM}}_{\pm} &= \frac{m \pm \sqrt{m^2-a^2 + D + a^2 k - D k }}{1 - k}, \\
    \text{For } \alpha = 0, \quad r^{\text{Dust}}_{\pm} &= \frac{2m +k\pm \sqrt{4 m^2-4 a^2 + 4 D + k^2 + 4 k m  }}{2}.
\end{aligned}
\label{eq:solutionslocationhorizon}
\end{equation}
The special case $D=0$ corresponds to the case where the turning point is at the horizon.  Apart from this case, any other case has $D> 0$ and we have
\begin{eqnarray}
    \alpha=\frac13 &:&\qquad r^{\text{Rad}}_+> r^{\text{Rad-H}}_{+}\ ,\qquad r^{\text{Rad}}_-< r^{\text{Rad-H}}_{-}\
\\
\alpha=-\frac13 &:&\qquad  r^{\text{DM}}_+ > r^{\text{DM-H}}_{+}\ ,\qquad r^{\text{DM}}_-< r^{\text{DM-H}}_{-}\
\\
\alpha=0 &:&\qquad  r^{\text{Dust}}_+> r^{\text{Dust-H}}_{+}\ ,\qquad r^{\text{Dust}}_-< r^{\text{Dust-H}}_{-}\
\end{eqnarray}

Thus we conclude that  $R(r)=0$ has only one solution
 outside the horizon. This proves the Lemma.

\subsubsection{Photon Region}

A photon orbit is characterized by $\dot r=0$ everywhere, i.e. $R=0$.
This condition is equivalent to $R=0$ at one point and $\frac{d \dot{r}}{d \lambda}=0$ everywhere. By differentiating $R(r)=\Sigma^2 \dot{r}^2$ we obtain
\begin{equation}
\begin{aligned}
	 \frac{1}{2 \Sigma^2} \frac{\partial R}{\partial r}=\frac{d \dot{r}}{d \lambda}+\frac{\dot{r}}{\Sigma} \frac{d \Sigma}{d \lambda} .
\end{aligned}    
\label{eq:partial-R}
\end{equation}
Now, taking into account that $\frac{d \Sigma}{d \lambda}  =\frac{\partial \Sigma}{\partial \theta}  \frac{d \theta}{d \lambda}+\frac{\partial \Sigma}{\partial r} \frac{d r}{d \lambda}  =\frac{\partial \Sigma}{\partial \theta} \dot{\theta}+\frac{\partial \Sigma}{\partial r} \dot{r}$, \eqref{eq:partial-R} takes the form
\begin{equation}
\begin{aligned}
 \frac{d \dot{r}}{d \lambda}  =-\frac{\dot{r}}{\Sigma}\left[\frac{\partial \Sigma}{\partial r} \dot{r}+\frac{\partial \Sigma}{\partial \theta} \dot{\theta}\right]+\frac{1}{2 \Sigma^2} \frac{\partial R}{\partial r} .
\end{aligned}  
\label{eq:partial-r-lambda}
\end{equation}
This equation represents the $r$-component of the geodesic equation for null geodesics.
If we set $\dot{r}=0 $ then  \eqref{eq:partial-r-lambda} becomes
$
	 \frac{d \dot{r}}{d \lambda}=\frac{1}{2 \Sigma^2} \frac{\partial R}{\partial r}.
$
Thus, a geodesic is a photon orbit if and only if
\begin{equation}
\begin{aligned}
	& R=0 \qquad \text{ at a single point,} \\
	& \frac{\partial R}{\partial r}=0 \qquad \text{everywhere .}
\end{aligned}    
\label{eq:photon-orbit-condition}
\end{equation}
Let us now 
assume  that there is a geodesic with $R=\frac{\partial R}{\partial r}=0$ at one point, that is $\dot{r}=0$ and $\frac{d \dot{r}}{d \lambda}=0$ at this point. This provides initial values and the unique solution is
$r=$ const., which represents a photon orbit.

In order to find the photon orbit's conditions, we start with 
\eqref{eq:R(r)-null-geodesics} describing radial geodesics
and set $R(r)=0$. We get
\begin{equation}
    \begin{aligned}
        &\Delta_k\left[(L-a E)^2+K\right]=\left[E\left(r^2+a^2\right)-a L\right]^2 ,
    \end{aligned}
    \label{eq:solving-photon-orbit-1}
\end{equation}
Defining $K_E=\frac{K}{E}$ and $L_E=\frac{L}{E}$, this can be  recast as
\begin{equation}
    K_E=\frac{\left[\left(r^2+a^2\right)-a L_E\right]^2}{\Delta_k}-\left(L_E-a\right)^2 .
\label{eq:solving-photon-orbit-2}
\end{equation}
Using the second photon orbit's condition,  $\frac{\partial R}{\partial r} =0$, we find

\begin{equation}
\begin{aligned}
	&  \frac{\partial R}{\partial r}=E^2\left(4 r^3+4 r a^2\right)-4 a L E r-\Delta_k^{\prime}\left[(L-a E)^2+K\right] = 0\\
	& \Rightarrow K_E=\frac{4 r}{\Delta_k^{\prime}}\left[\left(r^2+a^2\right)-a L_E\right]-\left(L_{E}-a\right)^2 \ ,
\end{aligned}   
\label{eq:partial-R=0}
\end{equation}
where $\Delta_k'\equiv \frac{\partial\Delta_k}{\partial r}$.
Equations \eqref{eq:solving-photon-orbit-2} and \eqref{eq:partial-R=0} form a system of two equations with two unknowns $L_E$ and $K_E$.
The solution is given by
\begin{equation}
    \begin{aligned}
 a L_E=\left(r^2+a^2\right)-\frac{4 r \Delta_k}{\Delta_k^{\prime}} .
    \end{aligned}
    \label{L_E-result}
\end{equation}
\begin{equation}    
{K}_E  =\frac{16 r^2 \Delta_k}{\left(\Delta_k^{\prime}\right)^2} -\left(L_E-a\right)^2 .
 \label{eq:K_E-final-result}
\end{equation}

Let us now consider the equation  \eqref{eq:theta-K} for 
$\Theta = \Sigma^2 \dot{\theta}^2$. This can be rearranged as

\begin{equation}
    \frac{\Sigma^2 \dot{\theta}^2 \; a^2 \sin^2 \theta  }{E^2} = \frac{K_E a^2 \sin^2 \theta}{E} - \cos^2 \theta \left[ (a L_E)^2 -a^4 \left( \frac{2 H}{E^2} + 1\right) \right]\geq 0\ .
    \label{eq:theta-K-final}
\end{equation}
Substituting \eqref{L_E-result}  and \eqref{eq:K_E-final-result} into the equation \eqref{eq:theta-K-final} we obtain 
\begin{equation}
    \frac{16 r^2 a^2 \sin^2 \theta \;\Delta_k}{E^2 (\Delta_k^\prime)^2} - \frac{1}{E^2}\left[ (r^2 + a^2) - \frac{4 r \Delta_k}{\Delta_k^\prime} -a^2 \right]^2 \sin^2 \theta -\cos^2 \theta \left[ (a L_E)^2 -a^4 (\frac{2H}{E^2} +1) \right]  \geq 0.
\end{equation}
At $\theta = \frac{\pi}{2}$ we find the inequality
\begin{equation}
    \begin{aligned}
         16  a^2 \Delta_k  \geqslant\left[r  \Delta_k^\prime-4  \Delta_k\right]^{2} ,
    \end{aligned}
    \label{eq:photon-region}
\end{equation}
which describes the photon region.

For the non-rotating case $a=0$, the equation \eqref{eq:photon-region} implies $r \Delta_k^{\prime}-4  \Delta_k=0$.
This means that the photon region degenerates into a photon sphere. 
In the Schwarzschild case, with $a=k=0$, $\Delta_k$ is replaced by  $\Delta_{\text{Schw}} = r^2 - 2 m r$, and the photon sphere consequently 
appears at $r= 3 m$, in agreement with the familiar result.

\subsection{Rotating Black Hole's Shadow}\label{sec:Kerr0-shadow}

To derive the shadow boundary curve on the observer's sky, it is necessary to determine the constants of motion for each light ray traced back from the observer's position. The spherical lightlike geodesic that serves as the asymptotic limit for these rays must share the same constants of motion, characterized by the radial coordinate \(r_p\). By allowing \(r_p\) to span its full range of values, from minimum to maximum and back, we obtain an analytical expression for the shadow boundary, represented as a curve parameterized by \(r_p\), in the form \((\psi(r_p), \zeta(r_p))\). The \((\psi(r_p), \zeta(r_p))\) are azimuthal and colatitude  angles in the observer's sphere, respectively.

First, we choose an appropriate set of orthonormal tetrads
\begin{equation}
\begin{gathered}
	e_0=\frac{\left(r^2+a^2\right) \partial_t+a \partial_\phi}{\sqrt{ \Sigma \Delta_k}}\quad , \qquad e_1=\frac{\partial_\theta}{\sqrt{ \Sigma}}, \\
	e_2=\frac{-\partial_\phi-a \sin ^2 \theta \partial_t}{\sqrt{ \Sigma}} \quad, \qquad e_3=-\sqrt{\frac{\Delta_k}{ \Sigma}} \partial_r .
\end{gathered}
\label{eq:tetrad}
\end{equation}
Next, we write down the equations representing the null rays in terms of $\partial_\mu$ where $\mu = \{r , \theta, \phi, t \}$ 
as
\begin{equation}
\begin{aligned}
    \dot{\lambda}=\dot{r} \partial_r+\dot{\theta} \partial_\theta+\dot{\phi} \partial_\phi+\dot{t} \partial_t ,
\end{aligned}
\label{eq:rays-partial-form}
\end{equation}
and
\begin{equation}
\dot{\lambda}=\alpha\left[-e_0+\sin \zeta \cos \psi\ e_1+\sin \zeta \sin \psi\ e_2\right.\left.+\cos \zeta\ e_3\right] .
\label{eq:rays-terad-form}
\end{equation}

The coefficients of $\partial_r$ and $\partial_\phi$ can be calculated from the expressions for $\dot r$ and $\dot \theta $. We find

\begin{equation}
    \begin{aligned}
        \dot{\lambda}=-\frac{a L-\left(r^2+a^2\right) E}{\sqrt{ \Sigma \Delta_k}}\left[\frac{1}{\sqrt{ \Sigma \Delta_k}}(a-\sqrt{\triangle} \sin \zeta \sin \psi)\left(\partial_\phi\right)\right.  \left.+\cos \zeta\left(\frac{\sqrt{\Delta_k}}{\sqrt{\Sigma}}\right) \partial_r+\cdots\right] .
    \end{aligned}
    \label{eq:rays-partialform-reexpress}
\end{equation}
By matching  the coefficients of $\partial_r$ in \eqref{eq:rays-partialform-reexpress} and \eqref{eq:rays-partial-form} we obtain
\begin{equation}
\begin{aligned}
	&  \cos \zeta=\frac{ \Sigma \dot{r}}{\left(r_0^2+a^2\right) E-a L} ,
\end{aligned}
\label{eq:Cos-zeta}
\end{equation}
where $r_0$ is the observer's position. Substituting $\dot{r}$ using \eqref{eq:radial-R}, we get
\begin{equation}
    \begin{aligned}
        &  \cos ^2 \zeta=1-\frac{\Delta_k K}{\left[E\left(r_0^2 +a^2\right)-a L\right]^2} \Rightarrow \quad \sin \zeta=\frac{\sqrt{\Delta_k K_E}}{\left(r_0^2+a^2-a L_E\right)} .
    \end{aligned}
    \label{eq:sin-zeta}
\end{equation}
Similarly, by matching the coefficients of 
$\partial_\phi$, we obtain
 \begin{equation}
    \begin{aligned}
	&  \sin \psi=\frac{L_E-a \sin ^2 \theta}{\sqrt{K_E} \sin ^2 \theta} ,
\end{aligned}
 \label{eq:sin-psi-3}
\end{equation}
where we used $K_E=\frac{K}{E}$ and $L_E=\frac{L}{E}$.

Note that the shadow is always symmetric with respect to the horizontal axis. In particular,  the points $(\psi, \zeta)$ and $(\pi-\psi, \zeta)$ correspond to the same constants of motion.

For $a>0$, the coordinate $\zeta$ takes its maximal value along the boundary curve at $\psi=-\frac{\pi}{2}$ and its minimal value at $\frac{\pi}{2}$.
We can easily derive the inequality determining the boundary curve by starting from \eqref{eq:sin-psi-3} and taking $\sin{\psi} = \pm 1$. Then using \eqref{eq:K_E-final-result} and \eqref{L_E-result}  one obtains
	
\begin{equation}
 \Sigma \Delta_k^{\prime}-4 r \Delta_k \mp 4 a r \sin ^2 \theta=0 .
\label{eq:boundary-curve-Kerr}
\end{equation}

The stereographic projection  maps the celestial sphere of the observer 
onto a plane that is tangent to this sphere at the pole at $\zeta=0$. In this plane we introduce (dimensionless) Cartesian coordinates:

\begin{equation}
\begin{aligned}
	& X\left(r_p\right)=-2 \tan \left(\frac{\zeta\left(r_p\right)}{2}\right) \sin \psi\left(r_p\right) ,\\
	& Y\left(r_p\right)=-2 \tan \left(\frac{\zeta\left(r_p\right)}{2}\right) \cos \psi\left(r_p\right),
\end{aligned}    
\label{eq:plot-shadow}
\end{equation}
where the upper half of the boundary curve is the mirror image of the lower half.

\subsubsection{The standard Kerr case}

As a check, we will now reproduce the known shadow of the
standard Kerr black hole.
We first  calculate the constants of motion $K_E$ and $L_E$ of spherical light ray at $r_P$ for the  Kerr metric with $\Delta_{\text{Kerr}} = r^2 = 2 m r +a^2$ instead of $\Delta_k$. For $K_E$, using  \eqref{eq:K_E-final-result}, we obtain

\begin{equation}
\begin{aligned}
	K^{\text{Kerr}}_E (r_p)=\frac{4 r_p^2\left(r_p^2-2 m r_p+a^2\right)}{\left(r_p-m\right)^2} .
\end{aligned}
\label{eq:K-E-rp}
\end{equation}
For $L_E$, using \eqref{L_E-result}, we get
\begin{equation}
\begin{aligned}
 a L^{\text{Kerr}}_E\left(r_p\right)=\frac{r_p^2\left(r_p-3 m\right)-r_p a^2-a^2 m}{r_p-m} ,
\end{aligned}
\label{eq:L-E-rp}
\end{equation}
where $\Delta_{\text{Kerr}} = r^2+a^2-2 m r$ has been used.

Finally, we can write \eqref{eq:sin-zeta} in terms of observer's position $r_0$ and spherical radial coordinate $r_p$ as
\begin{equation}
    \begin{aligned}
        	&  \sin^{\text{Kerr}} \zeta\left(r_p\right)=\frac{\sqrt{r_0^2+a^2-2 m r_0} \sqrt{K^{\text{Kerr}}_E\left(r_p\right)}}{r_0^2 +a^2 -a L^{\text{Kerr}}_E\left(r_p\right)} .
    \end{aligned}
    \label{eq:sin-zeta-2}
\end{equation}
Using \eqref{eq:K-E-rp} and\eqref{eq:L-E-rp} this becomes
\begin{equation}
    \begin{aligned}
       \sin^{\text{Kerr}} \zeta \left(r_p\right)=\frac{2 r_p \sqrt{r_p^2-2 m r_p+a^2} \sqrt{r_0^2-2 m r_0+a^2}}{r_0^2 r_p-r_0^2 m+r_p^3-3 r_p^2 m+2 r_p a^2} .
    \end{aligned}
    \label{eq:sin-zeta-rp}
\end{equation}
By using  \eqref{eq:sin-psi-3}, \eqref{eq:K-E-rp} and \eqref{eq:L-E-rp}, we can now write $\psi $
in terms of $r_p$ and $\theta_0$.  we obtain
\begin{equation}
 \sin^{\text{Kerr}} \psi\left(r_p\right)=-\frac{r_p^3-3 r_p^2 m+r_p a^2+a^2 m+a^2 \sin ^2 \theta_0\left(r_p-m\right)}{2 a r_p \sin \theta_0 \sqrt{r_p^2+a^2-2 m r_p}} .
 \label{eq:sin-psi-rp}
\end{equation}
Equations \eqref{eq:sin-zeta-rp} and \eqref{eq:sin-psi-rp} reproduce the known formulas determining the black hole's shadow for the Kerr case \cite{PERLICK2022}.

One can also recover  the Schwarzschild case, where $a=0$ and $r_p=3m$. This gives  
$$
\sin \zeta=\frac{3 \sqrt{3} m}{r_0}  \sqrt{1-\frac{2 m}{r_0}}
\ ,
$$ 
which is
the formula originally found  by Synge \cite{Synge}.

\subsection{Shadow for an observer at large distances: Rotating Kiselev Metric}

For the case of an observer sitting at large distances, we need to treat the shadow equations in the limit $r_0 \longrightarrow \infty$ or equivalently $r_0 \gg m$. First we take the distant observer's limit for equation \eqref{eq:sin-zeta-rp}. This gives 
\begin{equation}
\begin{aligned}
\lim _{r_0 \rightarrow \infty} \sin \zeta \left(r_p\right) 	\simeq \frac{\sqrt{K_E\left(r_p\right)}}{r_0} .
\end{aligned}
\label{eq:sin-zeta-distant-observer}
\end{equation}
The condition $m \ll r_0 $ implies $\frac{\sqrt{K_E}}{r_0}\ll  1$, therefore we can approximate $\sin{\zeta (r_p)} \sim \zeta\left(r_p\right)$. 
Then, taking into account \eqref{eq:sin-psi-rp}, $X(r_p)$ and $Y(r_p)$ in \eqref{eq:plot-shadow} can be approximated as

\begin{eqnarray}
\label{eq:x-plot-distant-observer}
	&&X\left(r_p\right)  \cong \frac{\alpha \sin ^2 \theta_0-L_E\left(r_p\right)}{r_0 \sin \theta_0} \ ,
    \\
  &&Y{\left(r_p\right)} \cong \mp \frac{1}{r_0} \sqrt{K_E\left(r_p\right)-\frac{\left(L_E\left(r_p\right)-a \sin ^2 \theta_0\right)^2}{\sin ^2 \theta_0}}  .
\label{eq:y-plot-distant-observer}
\end{eqnarray}
This provides an analytic representation of the Kiselev black hole shadow for a distant observer.

\vskip 1cm

\clearpage
\end{widetext}

\end{document}